\documentclass[pdflatex,sn-nature]{sn-jnl}

\usepackage{graphicx}%
\usepackage{multirow}%
\usepackage{amsthm}%
\usepackage{amssymb}
\usepackage{mathrsfs}%
\usepackage[title]{appendix}%
\usepackage{xcolor}%
\usepackage{textcomp}%
\usepackage{manyfoot}%
\usepackage{booktabs}%
\usepackage{algorithm}%
\usepackage{algorithmicx}%
\usepackage{algpseudocode}%
\usepackage{listings}%
\usepackage{glossaries}
\usepackage{gensymb}
\usepackage{hyperref}
\usepackage{wrapfig}
\usepackage{tablefootnote}
\usepackage{lineno}
\usepackage{cleveref}
\usepackage{caption}

\DeclareCaptionLabelSeparator{bar}{ | }
\crefname{figure}{Fig.}{Figs.}
\Crefname{figure}{Fig.}{Figs.}
\glsdisablehyper
\theoremstyle{thmstyleone}%
\theoremstyle{thmstyletwo}%
\theoremstyle{thmstylethree}%
\hypersetup{bookmarksdepth=2}

\def\mstrip{$M_{\rm strip}$}

\def\mcomp{$M_{2}$}

\def\countrate{ct\,s$^{-1}$}
\def\flux{erg\,s$^{-1}$\,cm$^{-2}$}
\def\kms{km\,s$^{-1}$}
\def\cms{cm\,s$^{-2}$}
\def\ergscm{erg\,s$^{-1}$\,cm$^{-2}$}
\def\ergs{erg\,s$^{-1}$}

\def\Lsun{L$_{\odot}$}
\def\Msun{M$_{\odot}$}

\def\msunyr{M$_{\odot}$\,yr$^{-1}$}

\def\Rsun{R$_{\odot}$}
\def\rsunyr{R$_{\odot}$\,yr$^{-1}$}

\def\HeII{He~\textsc{ii}}
\def\HeI{He~\textsc{i}}

\newcommand\arcsec{\mbox{$^{\prime\prime}$}}%
\newcommand\methods{(\nameref{Methods})}

\newcommand\method{\nameref{Methods}}
\newacronym[plural=SNe,firstplural=supernovae (SNe)]{SN}{SN}{Supernova}
\newacronym[plural=CMDs,firstplural=Color Magnitude Diagrams (CMDs)]{CMD}{CMD}{Color Magnitude Diagram}
\newacronym[plural=SEDs,firstplural=Spectral Energy Distribution (SEDs)]{SED}{SED}{Spectral Energy Distribution}
\newacronym{ZAMS}{ZAMS}{Zero Age Main Sequence}
\newacronym{MS}{MS}{main sequence}
\newacronym{AGB}{AGB}{Asymptotic Giant Branch}
\newacronym[plural=PNe,firstplural=Planetary Nebulae (PNe)]{PN}{PN}{Planetary Nebula}
\newacronym{WR}{WR}{Wolf Rayet}
\newacronym[plural=WDs,firstplural=White Dwarfs (WDs)]{WD}{WD}{White Dwarf}
\newacronym{NS}{NS}{Neutron Star}
\newacronym{UV}{UV}{ultraviolet}
\newacronym{PSF}{PSF}{Point-Spread Function}
\newacronym{FWHM}{FWHM}{Full Width Half Max}
\newacronym{SUMS}{SUMS}{Stripped star Ultraviolet Magellanic Clouds Survey}
\newacronym{MCPS}{MCPS}{Magellanic Clouds Photometric Survey}
\newacronym{SMC}{SMC}{Small Magellanic Cloud}
\newacronym[plural=RVs,firstplural=Radial Velocities]{RV}{RV}{radial velocity}
\newacronym{BHL}{BHL}{Bondi-Hoyle-Littleton}

\newcommand{\ellc}{\texttt{ellc}}

\newcommand{\Ultracam}{ULTRACAM}
\newcommand{\TheJoker}{\textit{The Joker}}
\newcommand{\XMM}{\textit{XMM-Newton}}
\newcommand{\Chandra}{\textit{Chandra}}

\def\MStrippedEvol{$3.38^{+4.34}_{-0.42}$}

\def\RStripped{$0.77^{+0.05}_{-0.27}$}
\def\Teff{$71^{+74}_{-7}$}

\def\LogG{$5.4^{+0.6}_{-0.5}$}
\def\XH{$0.10^{+0.25}_{-0.09}$}

\def\Luminosity{$4.15^{+0.84}_{-0.17}$}

\def\Period{$4.1693055(53)$}
\def\Amplitude{$151.4\pm2.5$}
\def\Voffset{$181.3 \pm 1.9$}
\def\Tzero{$58838.232974353$}
\def\fm{$0.062\pm0.003$}
\def\eccentricity{$e<0.045$}

\def\MCompanionOrbitMedian{1.38}
\def\MCompanionOrbitPercentiles{1.13--1.64}

\def\MCompanionOrbitDistPeak{$1.54$}
\def\MCompanionOrbitDistRange{1.04--2.08}
\def\MCompanionMin{$1.38 \pm 0.25$}
\def\MCompanionIncl{$1.54_{-0.50}^{+0.54}$} 
\def\MCompanionKDE{$1.58$}  
\def\MCompanionLC{$1.58_{-0.17}^{+0.56}$} 
\def\MCompanionHDIse{$1.40 - 2.41$}
\def\MCompanionHDInf{$1.33 - 4.51$}

\def\PWD{$0.6\%$}
\def\PBH{$3.4\%$}

\def\XMMFluxUnabsorbed{$<1.09\times 10^{-15}$ }

\def\XMMLum{$<4.79\times10^{32}$}

\def\ChandraCounts{$<9.3\times10^{-5}$}
\def\ChandraFluxUnabsorbed{$<1.8\times10^{-15}$}
\def\ChandraFluxAbsorbed{$<1.7\times10^{-15}$}
\def\ChandraLumUnAbsorbed{$<7.8\times10^{32}$}

\def\ChandraLumLog{$<32.9$}

\def\XBHLChandra{$1.55\times10^{33}$}

\defcitealias{Drout2023}{Drout \& G\"{o}tberg et al., 2023}

\begin{document}

\title[An Intermediate Mass Stripped Star in a 4.2-hour Orbit around a Compact Object]{An Intermediate Mass Stripped Star in a 4.2-hour Orbit around a Compact Object}

\author*[1]{\fnm{Bethany} \sur{Ludwig}}\email{bethany.ludwig@kuleuven.be}
\author[2]{\fnm{Maria R.} \sur{Drout}}
\author[4]{\fnm{Ylva} \sur{G{\"o}tberg}}
\author[2,3]{\fnm{Alexander} \sur{Laroche}}
\author[5,6]{\fnm{Katie} \sur{Auchettl}}
\author[7]{\fnm{Alejandro} \sur{Vigna--G\'omez}}
\author[8]{\fnm{Katelyn} \sur{Breivik}}
\author[1]{\fnm{Cole} \sur{Johnston}}
\author[1,9]{\fnm{Eva} \sur{Laplace}}
\author[10,11]{\fnm{Vikram S.}\sur{Dhillon}}
\author[1,9]{\fnm{Hugues} \sur{Sana}}

\affil*[1]{\orgdiv{Institute of Astronomy}, \orgname{KU Leuven}, \orgaddress{\street{Celestijnlaan 200D}, \postcode{3001}, \city{Leuven}, \country{Belgium}}} 

\affil[2]{\orgdiv{David A. Dunlap Department of Astronomy \& Astrophysics}, \orgname{University of Toronto}, \orgaddress{\street{50 St. George Street}, \city{Toronto}, \state{Ontario}, \postcode{M5S 3H4}, \country{Canada}}}

\affil[3]{\orgdiv{Dunlap Institute for Astronomy \& Astrophysics}, \orgname{University of Toronto}, \orgaddress{\street{50 St. George Street}, \city{Toronto}, \state{Ontario}, \postcode{M5S 3H4}, \country{Canada}}}

\affil[4]{
\orgname{Institute of Science and Technology Austria (ISTA)}, \orgaddress{\street{Am Campus 1}, \postcode{3400}, \city{Klosterneuburg}, 
\country{Austria}}}

\affil[5]{\orgdiv{School of Physics}, \orgname{The University of Melbourne}, \orgaddress{ 
\city{ Melbourne}, \postcode{VIC 3010}, \country{Australia}}}

\affil[6]{\orgname{OzGrav: The ARC Center of Excellence for Gravitational Wave Discovery}, \city{Melbourne}, \postcode{3010}, \country{Australia}}

\affil[7]{\orgdiv{Niels Bohr International Academy, Niels Bohr Institute}, \orgname{University of Copenhagen}, \orgaddress{\street{Blegdamsvej 17}, \city{ Copenhagen}, \postcode{2100}, \country{Denmark}}}

\affil[8]{
\orgdiv{McWilliams Center for Cosmology and Astrophysics}, 
\orgname{Carnegie Mellon University}, \orgaddress{  
\city{Pittsburgh}, \state{Pennsylvania},\postcode{15213}, \country{USA}}}

\affil[9]{\orgdiv{Leuven Gravity Institute}, \orgname{KU Leuven}, \orgaddress{\street{Celestijnenlaan 200D,  Box 2415}, \city{Leuven}, \postcode{3001}, \country{Belgium}}}

\affil[10]{\orgdiv{Astrophysics Research Cluster
School of Mathematical \& Physical Sciences}, \orgname{University of Sheffield}, \orgaddress{\street{Hounsfield Rd}, \city{Sheffield}, \postcode{S3 7RH}, \country{UK}}}

\affil[11]{\orgdiv{Instituto de Astrof\'{i}sica de Canarias}, \orgaddress{\postcode{E-38205} \city{La Laguna}, \country{Tenerife, Spain}}}

\def\name{\mbox{DGL-3}}

\abstract{\textbf{From their discovery as radio pulsars to landmark multi-messenger observations of their mergers, double neutron stars have transformed our understanding of the extreme Universe and the formation of heavy elements \citep{Drout2017,Abbott2017}. Yet, uncertainties in their formation result in predicted merger rates spanning orders of magnitude \citep{Tauris2017,mandel_rates_2022}.
The most critical stage occurs when a star is stripped of its hydrogen-rich envelope, causing the binary orbit to shrink drastically. 
Despite its fundamental role, the resulting tight binaries comprising a stripped star and neutron star have remained elusive.  Here we report a $\sim$3.4\,\Msun\ stripped star \citep{Ludwig2026,Drout2023,Gotberg2023} orbiting an unseen companion every 4.2 hours. 
Our orbital and spectroscopic analysis excludes a normal star, which would not fit within the orbit, as well as a second stripped star which would have been detectable in our spectra. Adding constraints from high cadence photometry gives a 99.4\% probability that the companion is too massive to be a white dwarf, making a neutron star the most likely companion. We model the system's future and predict that the stripped star will transfer mass again before exploding as a hydrogen-free supernova, leaving a bound pair of neutron stars that will merge after 220\,Myr. This system reveals a previously unconstrained phase of binary evolution, uncovering a new class of systems that probes the uncertain physics of massive stars, linking to observable gravitational-wave sources.
}}



\maketitle


\section{Main}\label{sec:intro}

Among the roughly 400 gravitational wave (GW) mergers detected to date \citep{LKV_Catalog_2026}, double neutron star (DNS) systems remain the only compact-binary class with a confirmed observation across the full electromagnetic spectrum. A single event, GW170817, which included the first kilonova detection, transformed our understanding of heavy-element nucleosynthesis \citep{Drout2017}, placed stringent constraints on the speed of gravity \citep{Abbott2017}, and gave a direct view of relativistic jet formation \citep{Mooley2018}. Despite this progress, and thorough theoretical work on binary \gls{NS} formation \citep{Tauris2017, VignaGomez2018, Ye2020}, the evolutionary pathways producing DNS mergers remain poorly constrained, as reflected in the wide spread of merger rates predicted by population synthesis models \citep{Mandel2022}. A key obstacle has been the absence of confirmed stellar progenitors immediately preceding compact neutron star binaries, leaving large uncertainties across this complex binary evolution pathway.

Current models predict that DNS binaries reach orbits tight enough to merge
through GW emission within the age of the Universe, or Hubble time, only after multiple phases of envelope-stripping \citep{Tauris2017}, including a common envelope
(CE) phase in which both stars become embedded within the donor's envelope.
Successful ejection of this envelope avoids a premature merger, resulting in a
tight binary composed of a NS and an envelope-stripped star.
To produce the second NS of a future DNS system, the stripped star must be massive enough to 
undergo core collapse  \citep{Tauris2015}, but not so massive that it forms a black hole (BH) instead. These ``intermediate mass'' stripped stars are expected to explode as hydrogen-poor supernovae \citep{Tauris2015}, which have low ejecta masses \citep{Drout2011} and sometimes small kicks \citep{Valli2025}, allowing the system to remain bound and create a tight-orbit DNS.
Identifying these immediate progenitors is therefore essential for constraining CE evolution and the formation efficiency of DNS systems.

However, until recently, intermediate-mass stripped stars capable of forming \glspl{NS} remained largely theoretical predictions \citep{Kippenhahn1967} without direct observations. While higher mass stripped star systems exist -- see for example  the Wolf-Rayet + compact object binary Cyg~X-3 \citep[e.g.,][]{vanKerkwijk1992,Zdziarski2013} -- they are  expected to form double BH or BH + NS binaries as opposed to DNS. At the explosion stage, a growing set of fast, faint supernovae with low ejecta masses ($\lesssim$0.5\,\Msun), are interpreted as stripped stars exploding in tight binaries with compact companions \citep[e.g.,][]{Drout2013, Tauris2013, De2018, Yao2020, Das2024, Moore2025}, but a NS or stripped star component has not been directly confirmed in any of them. This paradigm is now changing as a population of intermediate mass stripped stars was recently discovered in the Magellanic Clouds \citep{Drout2023, Ludwig2026, Gotberg2023}. 
Orbital solutions for five of these systems revealed unseen companions with
periods ranging from less than a day to nearly a year \citep{Laroche2026}, and
another stripped star in the Small Magellanic Cloud has recently been proposed
to orbit a compact object with a $\sim$20 hour period (Blomberg et al., submitted). Together, these detections suggest that the long-predicted progenitors of double neutron star systems are becoming observable, offering a way to calibrate binary evolution during this intermediate phase, between mass transfer and the final DNS merger, with direct observational constraints.

Here we present a detailed characterization of \name{}, a confirmed $\sim3.4$\Msun\ stripped star \citep{Drout2023} in an exceptionally tight 4.2 hour orbit with an unseen companion. We show that the companion is most likely a neutron star and model the binary's future evolution through mass transfer and core collapse, predicting the formation of a double neutron star that will merge through gravitational-wave emission within a Hubble time.
\name\ is thus the closest observed precursor of a DNS merger to date.

\bigskip
\noindent\textbf{\large The Current State of the System}
\bigskip

\noindent 
Originally discovered as a candidate stripped star based on the presence of excess ultraviolet light \citep[``SUMS\_b5d03";][]{Ludwig2026}, optical spectroscopy revealed a spectrum dominated by \HeII{} absorption lines confirming its status as a stripped star \citep[``Star 3";][]{Drout2023}. 
Spectroscopic and photometric fitting of the system with an extensive grid of stellar atmosphere models \citep{Gotberg2023} revealed a hot, compact, and hydrogen-poor star with a luminosity intermediate between known subdwarfs and Wolf-Rayet stars. 
The derived stellar properties (Table~\ref{tab:params}) closely match expectations for a core-helium burning stripped star
with an evolutionary mass of $\sim$3.4 \Msun\ derived
from the mass-luminosity relation of \citep{Gotberg2018} calibrated for stripped stars at the midpoint of core-helium burning. 
Notably, the spectroscopic fits of \cite{Gotberg2023} show no evidence for a luminous companion star in the optical spectrum of \name, requiring that any companion contribute $\lesssim$10\% of the optical flux, while a limited number of repeat spectroscopic observations presented in \cite{Drout2023} showed very large radial velocity shifts ($>$200 km s$^{-1}$), indicative of binarity. Together these restricted the set of viable companions to a faint \gls{MS} star, a second (fainter) stripped star, or a compact object.
Hereafter, we refer to the system as \name.

To determine the nature of \name{}'s unseen companion and its current binary configuration, we assembled a large data set consisting of: (i) 38 epochs of optical spectroscopy with Magellan/MagE \citep{Marshall2008} spanning 2019–2022, (ii) continuous \Ultracam\ imaging of \name\ with $\sim$20s exposures over a 3.4-hour period 
in the Super SDSS $u$, $g$, and $r$ filters 
(Extended Data \cref{fig:image-cutouts}), and (iii) 14 archival \XMM\ observations spanning  2001--2016, and 2 dedicated \Chandra\ ACIS-S pointings obtained in 2025 \methods. We summarize our results from the analysis of these data in Table \ref{tab:params}. 

\begin{figure}[p]
    \centering
    \includegraphics[]{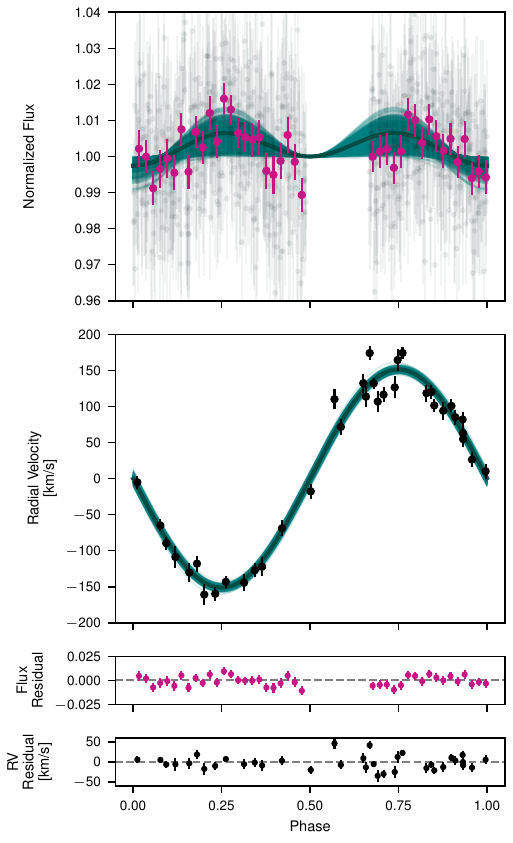}
    \caption{\textbf{The light curve and orbit of \name}. The light curve \textbf{(a)} and \gls{RV} curve \textbf{(b)} are plotted as a function of orbital phase, folded so that conjunctions occur at phases 0 and 0.5. Normalized and detrended ULTRACAM photometry is shown as the grey scatter points and binned in 5-minute intervals as the dark pink points. MagE radial velocities are shown as the black points. Thick teal lines show the best-fit \ellc{} model in \textbf{(a)} and the best-fit \emph{Joker} model in \textbf{(b)}. Thin teal curves show other models allowed by the data: \ellc{} solutions within $2\sigma$ of the best fit in \textbf{(a)}, and random draws from the \emph{Joker} MCMC posterior in \textbf{(b)}. The light curve shows double-peaked variability consistent with ellipsoidal modulation. The \gls{RV} curve shows large-amplitude variations from the short-period binary. \textbf{(c)} and \textbf{(d)} show the residuals of the binned photometry and radial velocities relative to the best-fit models.}
    \label{fig:ellcjoker}
\end{figure}

\textbf{The Binary Orbit:} We  measured radial velocities from the spectroscopic data \methods\, revealing a tight 
binary with a well-constrained orbital solution \methods, as shown in \cref{fig:ellcjoker} and listed in Table \ref{tab:params}. 
The preferred fit yields a very short period ($P$) of only $\sim4.2$ hours and is consistent with a circular orbit (\eccentricity; 95\% credible upper limit), which we adopt for subsequent analyses. 
The resulting semi-amplitude is large ($K$=\Amplitude{}~\kms), and the systemic velocity ($v_0$=\Voffset{}~\kms) confirms its membership to the \gls{SMC} \cite{ElYoussoufi2023}.

The orbital solution provides a measurement of the binary mass function, $f_m=P~K^3/(2 \pi G)=$\mcomp$^3\sin^3 i/($\mstrip$+$\mcomp$)^2$, where G is the gravitational constant, $i$ the orbital inclination, and  \mstrip\ and \mcomp, the masses of the stripped star and unseen companion, respectively. 
Adopting a distribution for the mass of the stripped star based on the evolutionary mass of \citep{Gotberg2023} (Table~\ref{tab:params}), we derive a minimum companion mass, reached for an edge-on orbit ($i=90$\degree), of $M_{2,\rm min}=$\MCompanionMin\Msun (median and 16/84th percentiles; see \method\ and Extended Data \cref{fig:minmass}).

Given the short period of the binary, our minimum mass measurements firmly establish that the companion cannot be a faint \gls{MS} star. Using MESA Isochrones and Stellar Tracks (MIST) evolutionary tracks \citep[version 1;][]{Dotter2016,Choi2016,Paxton2011,Paxton2013,Paxton2015}, we find that the radius of a zero-age \gls{MS} star, ${\rm R}_{\rm MS}$, with mass equal to the minimum companion mass 
always exceeds the Roche-lobe radius of the companion, ${\rm R}_{\rm RL, 2}$ \citep{Eggleton1983}, set by the constrained orbital configuration \methods. This is true for the entire range of possible primary star masses presented in \cite{Gotberg2023}, implying that a \gls{MS} companion would not physically fit within the orbit (see Extended Data \cref{fig:noms}). As we see no evidence that the stripped star is currently accreting substantial mass (e.g.\ strong emission lines or X-ray emission), a hydrogen-rich \gls{MS} companion is excluded.

While they are expected to be rare (Supplementary Information), double stripped star systems should exist \citep[e.g.,][]{Pols1994, Zapartas2026} and, unlike \gls{MS} stars, could fit within the binary orbit of \name{}. However, a stripped star companion would have been detectable in our data, as confirmed by creating and analysing artificial data sets  where representative model spectra of stripped stars have been injected in our multi-epoch data \methods.
We therefore conclude that the unseen companion to \name{} can neither be a \gls{MS} star,  nor another stripped star. \emph{The unseen companion  must therefore be a compact object.}

\textbf{The Light Curve:} To further characterize the system, we perform time-series photometry \methods{} to probe brightness variations associated with the binary orbit.
The continuous \Ultracam{} light curves cover $\approx83\%$ of the  orbital period and show coherent brightness variations of $\sim$2.5\% at twice the orbital frequency with maxima occurring near quadrature phases (i.e. 0.25 and 0.75; \cref{fig:ellcjoker}). This morphology is characteristic of ellipsoidal modulation, arising from the tidal deformation of the stripped star by a nearby companion.

For a given binary, the observed amplitude of ellipsoidal modulations depend heavily on the inclination angle of the system, deformability of the primary star (itself closely linked to primary's mass and radius), and mass of the secondary star. Quantitative assessment of this amplitude can therefore provide tighter constraints on these properties.
To this end, we model the ellipsoidal modulations from \name{} using the code \ellc{} and the observed $g$-band lightcurve, which has the highest signal-to-noise and lowest scatter \methods{}. We generate a large number of \ellc{} models, drawing from prior distributions for the stripped star mass/radius and observed inclination while holding other properties (i.e. the limb and gravity darkening coefficients) fixed. For each model the mass of the unseen companion is set by the combination of the stripped star mass, inclination, and our measured binary mass function. We then calculate the $\chi^2$ statistic and resulting likelihood between each model and the binned normalized light curve \methods{}.

\begin{figure*}[p]
    \centering
    \includegraphics[]{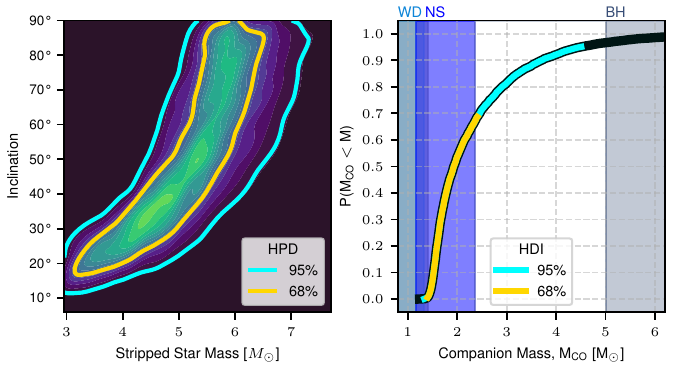}
    \caption{\textbf{Posterior from the likelihood-weighted light-curve grid.} Highest-density regions are 68\% (gold) and 95\% (cyan). \textbf{a}, joint kernel density of stripped star mass and inclination, with highest density contours. \textbf{b}, cumulative distribution of companion mass; thick segments mark the corresponding highest density intervals. Shaded bands show literature WD \citep{Caiazzo2021}, NS \citep{Martinez2015, Romani2022}, and BH \citep{Ozel2010} mass ranges.}
    \label{fig:kde}
\end{figure*}

Using these results, we generate probability distributions for the mass of the stripped star, the mass of the unseen compact object companion, and the orbital inclination of the system, weighting each model by its likelihood. We find that if the stripped star in \name{} is on the higher (lower) end of the allowed distribution, we would require a higher (lower) inclination angle to reproduce the observed variability amplitude, and that very low inclination angles are generally disfavored (\cref{fig:kde}, \textbf{a}).  The corresponding companion mass distribution (\cref{fig:kde}, \textbf{b})  only has \PWD\ of the posterior below the Chandrasekhar limit of $1.4$\,\Msun\ which makes a WD companion unlikely. 
The marginal posterior peaks at \MCompanionKDE~\Msun\ (hence, in the \gls{NS} range). The 68\%\ Highest Density Interval (HDI) extends  over [\MCompanionHDIse]~\Msun. 

Only \PBH\ of the posterior lies above the floor of dynamically confirmed BHs \citep{Ozel2010}; the remainder of the high-mass tail falls in the NS/BH mass gap. Overall, between our derived companion mass distribution and the current measured masses of Galactic double neutron stars \citep{Chattaraj2026}, \emph{a neutron star is the most probable companion.}

\textbf{The X-rays:} Given that the system is a stripped star with a likely NS companion, we explored the possibility of X-ray emission due to wind accretion onto a compact object. 
No significant source is detected at the optical position in any individual or stacked archival
\XMM\  data, nor in the targeted Chandra data we acquired.
For the latter, we find a 99.87\% 
upper limit on the unabsorbed X-ray luminosity in the 2--7 keV band to be  \ChandraLumUnAbsorbed\ \ergs (see Table~\ref{tab:params} and \nameref{Methods}). 

To place this upper limit in context, we use the Bondi--Hoyle--Lyttleton (BHL) formalism \citep{BondiHoyle1944} to model wind accretion onto a 1.4 \Msun\ NS. Adopting the spectroscopic and orbital parameters of \name{} along with a measured upper limit on the wind mass-loss of $\dot{M}_{\rm wind}=10^{-9}$\,\Msun\,yr$^{-1}$ \citep{Gotberg2023} along with standard efficiency factors \methods{}, the predicted X-ray luminosity is $L_{2-7\mathrm{keV,BHL}}\sim$\XBHLChandra\ergs. This is marginally above our  \Chandra\ limit, implying the source could sit near the detection threshold. However, the prediction is highly sensitive to poorly constrained wind properties; either lowering the assumed mass loss rate or increasing the wind speed would place the predicted flux below our detection threshold \methods{}. In addition, in low wind regimes, the system may occupy the propeller regime, in which the magnetosphere prevents material from reaching the surface and quenches X-ray emission \citep{Illarionov1975}. Our \Chandra\ upper limits are thus entirely consistent with a neutron star companion; intermediate-mass stripped star winds are likely too weak to produce a bright X-ray source in the Magellanic Clouds.

\bigskip
\noindent\textbf{\large Forming a merging double neutron star}
\bigskip

\noindent \name{} allows us to model the evolution of a stage in the formation of merging neutron stars that was previously observationally unconstrained. By anchoring our calculations to the measured properties of this exceptionally tight binary, we explore its future evolution using the binary stellar evolution code MESA \citep{Paxton2011, Paxton2013, Paxton2015, Paxton2018, Paxton2019, Jermyn2023} \methods{}. We evolve the current system through subsequent mass transfer and supernova, and then integrate the binary orbit until the system merges through gravitational wave emission.

\begin{figure*}
\centering
\includegraphics[width=\textwidth]{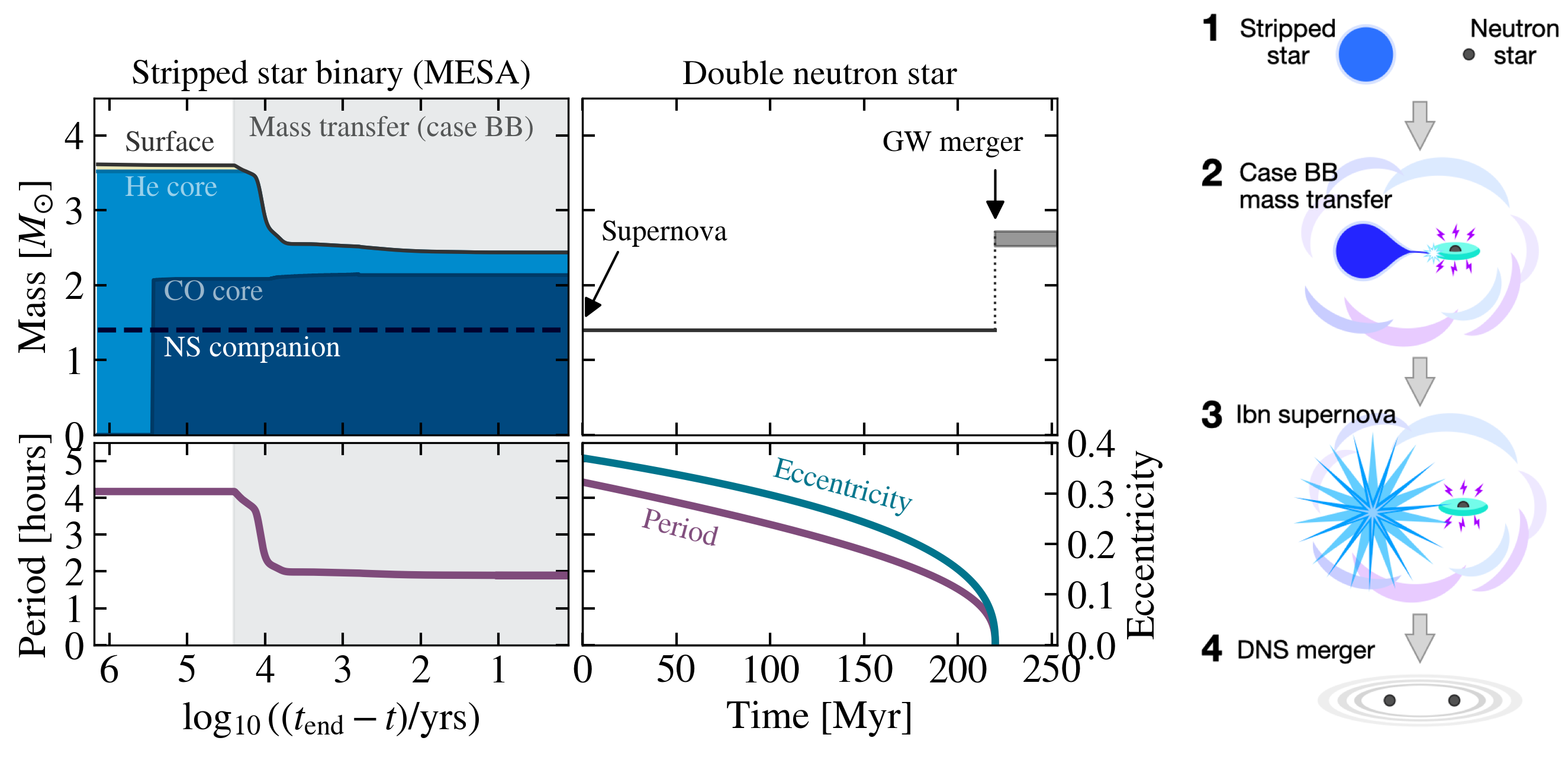}
\caption{\textbf{Evolution from stripped star binary to merging double neutron star.} Mass and orbital evolution of the stripped star binary model before (\textbf{a}) and after (\textbf{b}) the supernova. \textbf{a,} \textit{Top:} masses of the stripped star, its remnant, and the companion. The surface is marked by a line; the interior is colored by the leftover hydrogen layer (yellow), the helium core (light blue) and the carbon-oxygen core (dark blue).
Grey shading indicates the phase of Roche-lobe overflow referred to as case BB mass transfer and during which all hydrogen is lost. 
\textit{Bottom:} orbital period evolution, with circular orbits assumed until core collapse. 1\Msun{} of material is lost in the helium-rich supernova, which makes the resulting double neutron star orbit eccentric. Gravitational-wave radiation then shrinks and circularizes the orbit, producing a neutron-star merger after $\sim 220\,\mathrm{Myr}$. 
\textbf{b,} Predicted future evolutionary sequence of the observed system. The stripped star fills its Roche lobe during helium-shell burning (case BB mass transfer), explodes as a stripped-envelope supernova, and leaves a bound neutron-star binary that merges through gravitational-wave emission. }
\label{fig:model}
\end{figure*}

We place a $3.6$\,\Msun\ stripped star structure model, custom-made to reproduce the observed properties of the present day system (Table~\ref{tab:params};  Supplementary Information), in a circular 4.2~hour orbit with a $1.4$\,\Msun{} point-mass companion and evolve the system forwards in time, as visualized in Fig.~\ref{fig:model}. The central helium burning lasts 1.4\,Myr, after which helium ignites in a shell, causing the stripped star to expand. Because the present-day orbit is so tight, an expansion of only 0.5\,\Rsun\ is needed for the star to fill its Roche Lobe and initiate a new phase of mass transfer, commonly referred to as Case BB \citep[e.g.,][]{Dewi2003,VignaGomez2018}, which 
lasts 26\,kyr.
To resemble a NS fed through Roche lobe overflow, we limit accretion to the Eddington accretion rate ($\sim 10^{-8} M_\odot$/yr) and eject the remaining material with the orbital angular momentum of the NS \citep{soberman_stability_1997}. 
This highly non-conservative mass transfer removes 1.2\,\Msun{} of material from the system and reduces the orbital period from 4.2 to 1.9 hours (\cref{fig:model}), producing an even more compact binary immediately preceding the second supernova. 

During case BB mass transfer, the system should become a bright X-ray binary although the NS only accretes 0.00072\,\Msun{} of the transferred material. We estimate that the accretion spins it up to 79 ms \citep[Eq.~14 of][]{Tauris2012}, thus potentially becoming a mildly recycled pulsar. Assuming that the material that is lost from the system leaves with the orbital velocity of the NS ($\sim 500$\,km/s) and is distributed spherically around the system, it should pile up in a circumstellar medium structure that is densest close to the system ($\sim 10^{-12}$\,g/cm$^3$ at the orbital separation) and decreasing outwards ($\sim 10^{-25}$g/cm$^3$ at 1 pc) \citep[][Supplementary Information]{WeiDandan2026}. These shells of material are mostly composed of helium, and  are expected
to be visible during the bright X-ray binary stage.  

The model develops a metal core of 2.1\,\Msun\ (\cref{fig:model}), which is well above what is necessary to reach core collapse \citep[$>1.45$\,\Msun{},][]{Tauris2013}. At explosion, the star has lost all hydrogen and still fills its Roche lobe. The absence of hydrogen along with the remaining 0.2\,\Msun{} of helium suggests that the star will explode as a stripped-envelope supernova. Given the dense circumstellar material, the explosion would likely appear 
as a type Ibn \citep{Hachinger2012, Sollerman2020, Dessart2022}. 
Assuming the formation of a 1.4\,\Msun\ NS, we predict 
1\,\Msun\ of material would be ejected from the binary, with a mixed composition of helium, carbon, oxygen, neon, magnesium and silicon.

At core collapse, the orbital velocity is nearly 800\,\kms, meaning that an unusually large neutron star kick would be needed to disrupt the system. 
Furthermore, at least a subset of stripped stars should experience small to no natal kicks given the configurations of Be X-ray binaries \citep{Valli2025}, which are direct descendants of Be + stripped star systems. We therefore adopt a symmetric explosion with no natal kick.
We find that the kick originating from the instantaneous 1\,\Msun{} loss is not sufficient to disrupt the binary, thereby producing
a bound DNS system with an eccentricity of 0.37 and an orbital period of 4.4 hours. 
The binary then radiates gravitational waves at such high rates that the NSs merge after only 220 Myr (\cref{fig:model}, \cite{Peters1964}). This is much shorter than a Hubble time,
demonstrating that \emph{the observed configuration of \name{} can lead directly to a merging double neutron star.}

The predicted formation of a DNS merger
is robust across a range of evolutionary assumptions \methods. The largest uncertainty arises during the Case BB mass transfer phase, because of how poorly understood angular momentum loss is during non-conservative mass transfer \citep[e.g.,][]{Kovacevic2026}. If, for example, such mass transfer forms a circumbinary disk, it could drain angular momentum so efficiently that the stars merge before the second supernova. While such a merger likely would result in an interesting blue transient, the angular momentum loss model we adopt is standard. We also model the system using weaker angular momentum loss, and different component masses, but find a DNS merger in every scenario.
We conclude that \name{} therefore provides an exceptional observational benchmark for our binary evolution models that is especially useful for late-stage, interacting, massive binaries.

\bigskip
\noindent\textbf{\large From a Benchmark to Binary Evolution}
\bigskip

The expected number of stripped stars with NS companions remains highly uncertain, ranging from a few to several hundred in the Milky Way, and from fewer than one to several tens in the Magellanic Clouds \citep{Gotberg2020B, HovisAfflerbach2025, Martinez2025,
Zapartas2026}. This broad range reflects uncertainties in binary interaction physics that propagate directly into DNS formation and merger rates. As such, these systems offer a unique opportunity to constrain mass-transfer processes in massive binaries, particularly during the shortest and most dynamically complex evolutionary phases. Recent detections of stripped stars with unseen companions \citep[][Blomberg et al., submitted]{Laroche2026} suggest that further pre-DNS binaries may soon be uncovered, each offering a direct test of NS formation and merger pathways.

\name{} shows that such a system can be found and characterized in detail, making it the first benchmark at a critical stage of DNS formation. Our forward modeling shows consistency with both the Galactic population of millisecond
pulsars and the DNS merger GW170817 \citep{VignaGomez2018}. Its 
properties also tie it to other gravitational-wave sources: were \name{} within $\sim$100--1000\,pc, it would be detectable by low-frequency gravitational-wave observatories such as LISA \citep{Gotberg2020B}.

Future work can sharpen this benchmark from both the theoretical and the observational side. On the modeling side, our one-dimensional models of the future evolution of \name{}  can be extended with three-dimensional hydrodynamic simulations. These can directly follow how angular momentum is lost from the orbit, and so further refine
the predicted orbital outcome, the effect of the residual surface hydrogen, and the X-ray and GW emission. The formation of the present system has also not yet been explored: binary-evolution and population synthesis models can test the implications of our measured stellar and orbital properties on the efficiency and depth of stripping achieved in the common envelope phase.
On the observational side, measurements of the
star's rotation and possibly its pulsations would show whether it is tidally synchronized and how it rotates internally---providing important constraints on angular momentum transport in massive binary stars.
The system thus connects the observed properties of a stripped star to several key outcomes of binary evolution, including X-ray binaries, interacting supernovae, pulsars, and merging DNSs.

\begin{table}[p]
\centering
\caption{\textbf{Binary parameters.} System, stripped-star, orbital, and X-ray upper-limit values for DGL-3. Unless stated otherwise, uncertainties are 68\% credible intervals. For symmetric posteriors we report the median with the 16th--84th percentile range; for asymmetric posteriors we report the mode with the 68\% highest-density interval (HDI). Literature values are quoted at their original $1\sigma$.}
\label{tab:params}
\begin{tabular}{c||c}
\toprule 
\multicolumn{2}{c}{Parameters from Previous Work}  \\ 
\midrule
\emph{System Properties} \citep{Ludwig2026, Drout2023}
& \\ [4pt]
Right ascension (J2000) &  0$^\mathrm{h}$ 57$^\mathrm{m}$ 40$^\mathrm{s}$.09 \\[2pt] 
Declination (J2000) & -71\degree\ 59' 16".5  \\ [4pt]
\emph{Stripped Star} \citep{Gotberg2023} 
& \\ [4pt]
Evol. Mass, M$_{\rm evol}$  
&\MStrippedEvol\; \Msun  \\ [2pt]
Effective Radius, R$_{\rm eff}$  & \RStripped\; \Rsun  \\ [2pt]
Luminosity, $\log_{10}{\rm L_{Bol}}$  & \Luminosity\; \Lsun \\ [2pt]
Temperature, T$_{\rm eff}$ & \Teff\; kK \\ [2pt]
Surface Gravity, $\log_{10}{\rm g}$ & \LogG\; \cms \\ [2pt]
Surface Hydrogen Mass Fraction, $X_{\rm H, \, surface}$ & \XH\; \\ [2pt]
\midrule
\multicolumn{2}{c}{Orbital Configuration}\\
\midrule
Orbital Period, $P$  & \Period\ hr \\ [2pt]
Eccentricity, $e$\footnotemark &$<0.045$ \\  [2pt]
Semi-Amplitude, $K$ & \Amplitude\ \kms   \\ [2pt]
Barycentric Velocity, $v_0$ &  \Voffset\ \kms \\ [2pt]
Time at phase = 0, $T_0$ & \Tzero\,MJD \\ [2pt]
Mass function, $f(m)$ &  \fm\ \Msun\\[4pt]
\midrule
\multicolumn{2}{c}{Companion Mass Constraints}\\
\midrule 
Edge-on ($i=90^\circ$), $M_{2,\min}$  & \MCompanionMin\Msun \\  [3pt]
Isotropic inclination, $\langle M_{2, \rm inc}\rangle$ & \MCompanionIncl\Msun \\  [3pt]
Light-curve inclination, $M_{2, i_{\mathrm{LC}}}$ & \MCompanionLC\Msun \\  [3pt]
\midrule
\multicolumn{2}{c}{\Chandra\ (2-7 keV) Upper Limits }\\
\midrule 
Count Rate  &  \ChandraCounts\ \countrate \\[2pt]
Absorbed Flux  & \ChandraFluxAbsorbed\ \flux \\[2pt]
Unabsorbed Flux & \ChandraFluxUnabsorbed\ \flux \\ [2pt]
Log Luminosity  & \ChandraLumLog\ \ergs \\ 
\midrule
\bottomrule
\end{tabular}
\footnotetext[1]{The eccentricity is consistent with zero. We quote the $95\%$ credible upper limit from the orbital fit, but fix $e = 0$ in all subsequent analysis.}
\end{table}

\clearpage
\section{Methods}\label{Methods}
This section describes the spectroscopic, photometric, and X-ray observations, as well as the data analysis and binary evolution models used in this work. Additional figures and tables supporting the analysis are provided as Extended Data. 

\paragraph{Spectroscopic Observations}

We obtained 38 epochs of optical spectra  with the Magellan Echellette spectrograph \citep[MagE;][]{Marshall2008} on the 6.5 meter Magellan Baade telescope at Las Campanas Observatory in Chile. The spectra cover the wavelength range from $\sim$3395\AA\ to 7855\AA\ with a spectral resolving power of $\mathcal{R} \sim 4100$, as defined by the entrance slit width of 0.85\arcsec. Data reduction followed the procedure outlined in \cite{Drout2023,Laroche2026}, with one modification: due to the short orbital period and large radial velocity shifts (derived below) we do not stack multiple exposures taken sequentially on a given night. Rather we analyze each individual (20min) exposure independently.

\paragraph{RV Measurements}
We adopt a line fitting approach to measure the radial velocity shifts of each epoch spectrum \cite[cf.][\nameref{Methods}]{Sana2013}. Given the low signal-to-noise ratio of our epoch spectra (S/N$\sim$10--15; due to analyzing each 20 min exposure independently), we opt for line fitting over cross-correlation, which can perform poorly at this S/N \citep{Massey2023}. We summarize our method below and refer to \citep{Laroche2026} for a complete description.

We measure RVs from seven photospheric absorption lines of the stripped star: H$\delta$, He~\textsc{ii}\,4200, H$\gamma$, He~\textsc{ii}\,4542, He~\textsc{ii}\,4686, H$\beta$ and He~\textsc{ii}\,5412.\footnote{Each Balmer line is blended with a He~\textsc{ii} Pickering ($n\to4$) line.} 
To build line templates, we first fit Voigt profiles to each line in every epoch ($\pm$15\,\AA\ around line center for Balmer lines, $\pm$8\,\AA\ for He~\textsc{ii}). We then shift and add the lines from all epochs using  preliminary RV estimates to produce a high-S/N stack, and fit a Voigt profile to the stack to obtain the best-fit template of each line. 
For each epoch, we shift all templates simultaneously and minimize $\chi^2$ to obtain the best-fit RV. 
To estimate the uncertainty, we repeat this fit on 1000 Monte Carlo realizations of the epoch spectrum, perturbing the flux in each pixel with Gaussian noise drawn from its measured error. We fit a Gaussian to the resulting RV distribution, taking its mean as the epoch RV and its standard deviation as the 1$\sigma$ uncertainty (typically 10\,km\,s$^{-1}$).
All of our measured RVs are presented in Extended Data  Table~\ref{tab:RV_measurements}. Times are HJD on the Barycentric Coordinate Time (TCB) scale. Two RV measurements are rejected due to poor data quality.

\paragraph{Orbital Parameters}
To determine the orbital parameters, we use \TheJoker\
 \cite{Joker12017,Joker22017} to fit a Keplerian model with period $P$, semi-amplitude $K$, eccentricity $e$, argument of periastron $\omega$, mean anomaly at the reference epoch $\phi_0$ (which sets the phase of pericenter passage), and systemic velocity $v_o$.
 The Joker follows rejection sampling on a wide prior distribution of orbital parameters with MCMC. For the rejection sampling, we randomly try many possible orbits (20 runs of $10^6$ trials). The period is log-uniform, $p(P)\propto 1/P$, from 0.1 to 100 days. The eccentricity follows the Beta distribution from \citep{Kipping2013}. The size of the star’s velocity wobble ($\sigma_{K_0}$) and the system’s overall speed ($\sigma_{v}$) are each allowed to be large: both use a scale of $100$\,\kms. MCMC is then initialized from a surviving sample: 4 chains, 4,000 tuning steps, and 12,500 retained draws per chain (50,000 samples). With the period already tightly constrained by the rejection sampling, the MCMC period prior is restricted to 0.1--1 day.

For near-circular orbits, the argument of pericenter $\omega$ and the pericenter phase $\phi_0$ are degenerate, because no pericenter is defined at $e=0$. We therefore sample the well-constrained combination $\phi_0+\omega$ to improve MCMC mixing.
The eccentric posterior is consistent with a circular orbit (median e=0.016). As the eccentricity is bounded at zero, we report a one-sided 95\% credible upper limit from the MCMC samples, \eccentricity, defined as the 95th percentile of the eccentricity chain (see Extended Data \cref{fig:eccentriccorner}). The mean eccentric orbit has $\chi^2=82.85$ with 31 degrees of freedom (dof). Fixing $e=0$ and solving for a circular orbit yields $\chi^2=83.19$ with 32 dof. An F-test of this nested comparison gives $F=0.127$, well below $F_{\rm crit}=4.160$ at $95\%$ confidence. The extra eccentricity parameter does not significantly improve the fit, and we adopt the circular solution shown in Extended Data \cref{fig:circularcorner}. 

The posterior median of each orbital parameter, with a 68\% equal-tailed (16th-84th percentile) credible interval, is listed in Table~\ref{tab:params}. For the roughly Gaussian posteriors of $P$, $K$, and $v_o$ this interval is equivalent to $1\sigma$.

\paragraph{Orbital Companion Mass Constraints}

To place initial constraints on the mass of the unseen companion star in \name{} we combine the binary mass function derived from the orbital solution with measurements of the mass of the stripped star primary from \cite{Gotberg2023}. 
Here we adopt the evolutionary mass  as the latter as it is more robust than the spectroscopic mass \cite{Gotberg2023} and the derived stellar properties of \name{} closely match expectations for a core-helium burning stripped star. In addition, the evolutionary mass  
is derived from a luminosity that varies only modestly across the helium-burning lifetime, making it insensitive to the star's precise evolutionary state. However $M_{\rm evol}$ has strongly asymmetric error bars, which are much larger above the best-fit value than below it. We therefore follow the procedure outlined in \cite{Laroche2026} and sample the stripped star mass from a truncated split-normal distribution centered on the best-fit evolutionary mass, with independent left and right widths set by the $1\sigma$ uncertainties and truncated to the corresponding interval ($2.96-7.72$\,\Msun). 

Sampling from both this distribution for $M_{\rm{strip}}$ and the posterior for the binary mass function, $f_m$, and fixing the inclination angle to $i=90$\degree\, we find the minimum companion mass distribution shown in Extended Data \cref{fig:minmass}a. This distribution has a median value of \MCompanionOrbitMedian\,\Msun\ and a 68\% confidence interval (as defined by the 16th and 84th percentiles) of 
\MCompanionOrbitPercentiles\,\Msun. 

If we instead sample inclination angles from an isotropic distribution, we find the distribution of possible companion masses shown in Extended Data \cref{fig:minmass}b. This distribution peaks at \MCompanionOrbitDistPeak\,\Msun\ and has a 68\% highest-density interval of \MCompanionOrbitDistRange\,\Msun. Every individual companion mass draw exceeds the corresponding minimum mass, but the lower bound of this interval falls below that of the minimum-mass interval. This is because the inclination-sampled distribution rises sharply at the minimum mass (edge-on) and has an extended tail at higher masses (low inclinations). The narrowest interval containing 68\%\ of the samples is pulled toward the low-mass edge. The lower bound therefore reaches into the low-mass tail of the minimum-mass distribution, from draws with low $M_{\rm strip}$, whereas the 16th percentile used for the minimum mass excludes that tail.

\paragraph{Companion Size Constraints from Orbit}
We combine the orbital solution derived above with stellar evolution models to place constraints on the type of companion star that could physically fit within the binary orbit. To assess the possibility of faint, low mass, main sequence stars we use the [Fe/H]$=$$-$0.85 and $v/v_\mathrm{crit} = 0.4$ models from MIST \citep[version 1;][]{Dotter2016,Choi2016,Paxton2011,Paxton2013,Paxton2015}. For a given (assumed) stripped star primary mass we then perform the following procedure: (i) we determine the minimum mass of unseen companion based on our derived binary mass function, (ii) we calculate the size of the Roche radius, $R_{\rm RL, 2}$, for this minimum mass companion following \citep{Eggleton1983}, and (iii) we compare the size of $R_{\rm RL, 2}$ to the radius of a MIST zero-age main sequence (ZAMS) with a mass equal to the minimum companion mass. We carry out this procedure for the full range of primary stripped star masses allowed within 1$\sigma$ for \emph{both} the evolutionary and spectroscopic masses in \cite{Gotberg2023}. In all cases the radius of the ZAMS star exceeds $R_{\rm RL, 2}$, implying it would not fit (see Extended Data \cref{fig:noms}). As both pre-MS stars and more evolved MS stars have larger radii than ZAMS stars, they would also be excluded. In contrast, repeating this process using the stripped star models of \cite{Gotberg2018} (which have radii a factor of $\sim$2-3 smaller than MS stars of similar mass) we find that they would fit within the binary orbit (see however next paragraph).

\paragraph{Spectral Disentangling}
We assessed whether a stripped star companion could be recovered
from our multi-epoch spectroscopy using \HeII~$\lambda$4686, one of the strongest features of intermediate mass stripped stars.  We applied a shift-and-add spectral disentangling code \citep{Gonzalez2006,Shenar2020, Shenar2022} to spectra at MagE resolution
($R \approx 4100$) in two controlled tests: (i)~injection of synthetic companions into the observed spectra, and (ii)~a simulated double stripped star binary. 

\textit{Model setup.} 
In both tests, the primary is the best-fit synthetic
model for \name\ from \citep{Gotberg2023}, and the companion is one of six
stripped-star models from \citep{Drout2023} with masses between 1.05 and
2.04\,\Msun. All models were degraded to MagE resolution and scaled to the
distance of the SMC. We define the companion's fractional flux contribution as
$\ell_2 = F_2/(F_1+F_2)$, measured in the continuum window 4700--4725\AA\ near
the \HeII\ line. For the chosen companions, $\ell_2$ corresponds to 13--35\% of
the total optical continuum. For the primary, we adopt our measured
semi-amplitude, $K_1 = 151.4$\kms, and the best-fit evolutionary mass from
\citep{Gotberg2023}, $M_1 = 3.38$\,\Msun. The companion's semi-amplitude follows
from momentum conservation, $K_2 = K_1 M_1 / M_2$.

\textit{Injection test.} We injected each companion model into the observed
spectra as
\begin{equation}
  F = (1-\ell_2)\,F_{\mathrm{obs}} + \ell_2\,F_2(v_2),
\end{equation}
where $F_2(v_2)$ is the companion model Doppler-shifted to its expected orbital
velocity at each epoch. 

\textit{Simulated binary.} For the fully simulated case, we Doppler-shifted the
primary and companion models to the velocities expected for a circular
double-lined spectroscopic binary (SB2) at the observed epochs and combined
them. We then added Gaussian noise using the observed per-epoch MagE flux
errors.

For each set of spectra, we ran the disentangling on
the \HeII~$\lambda$4686 line, initializing the companion flux contribution at
the injected value of $\ell_2$ and searching a dense grid of $K_2$ values
spanning 0.4--1.8 times the dynamical estimate.
We find that \HeII~$\lambda$4686 recovers $K_2$ to within
$\sim\!25\%$ for five of the six companion masses in both tests, with
a larger bias only at the lowest-mass end of the composite grid, and
that the disentangled companion line exceeds $\mathrm{S/N} > 3$ at
every mass (Extended Data \cref{fig:k2-recovery-snr}). Here the line S/N is the
absorption depth in the disentangled secondary spectrum~$\mathrm{B}$
divided by the continuum RMS in the line-window edges.

To visualize where such a companion would appear in the data, we
constructed mean-subtracted dynamical spectra of \HeII~$\lambda$4686
from the MagE observations and from the simulated binaries on a shared
velocity--phase grid ($|\Delta v| \le 600$\kms,
$10$\kms\ bins, two orbital cycles; Extended Data \cref{fig:dynamical-4686}). Overlaid primary and companion RV
curves ($v_1$, $v_2$) mark the expected loci of the secondary
signature for each companion mass.

We repeated the same injection and simulated-binary tests for
additional optical lines individually and in combination---\HeII\
$\lambda$5412, \HeII/\textrm{H}$\beta$~$\lambda$4860, \HeI\
$\lambda$4471, and \HeI~$\lambda$5876. Among these, \HeII\
$\lambda$4686 provides the most consistent $K_2$ recovery together
with companion-line $\mathrm{S/N} > 3$ across the full mass grid.
Applying the same disentangling procedure to the observed (uninjected)
MagE time series of \name\ yields no convincing signature of a second
stripped star.

\paragraph{Ultracam Images}
We observed \name\ with the high-speed imager \Ultracam\ \citep{Dhillon2007}
mounted on the 3.5\,meter New Technology Telescope (NTT) at La Silla Observatory
on December 18, 2022. 
The observations were obtained continuously over a single $\sim$3.4-hour
baseline, yielding 617 frames. The exposure time was 19.985\,s for 616 frames,
with 0.024\,s of dead time between exposures (a cadence of 20.009\,s). The first
frame was longer, at 20.4\,s.
\Ultracam\ provides simultaneous imaging in three photometric bands; we used
the Super SDSS $u$, $g$ and $r$ filters described in \citep{Dhillon2021}.

All frames were reduced with the \textsc{hipercam} data reduction package,\footnote{\url{https://github.com/HiPERCAM/hipercam}} which supports the HiPERCAM, \Ultracam, and ULTRASPEC high-speed cameras. The reduction included bias subtraction, flat-field correction, and dark current removal. \cref{fig:image-cutouts} shows $20\arcsec\times20\arcsec$ cutouts centered on \name, constructed from a stack of the first 30 frames ($\sim$10 minutes of total exposure) in each band. \name\ becomes markedly brighter relative to its neighbors at shorter wavelengths, as expected for an intermediate-mass stripped star, which is hot and intrinsically blue.

\paragraph{Ultracam Photometry}
We first performed relative aperture photometry on the \Ultracam\ images using the \textsc{hipercam} reduction pipeline. Although the pipeline supports aperture radii that scale with the \gls{FWHM} of the \gls{PSF}, the resulting light curves showed non-negligible brightness variations correlated with seeing, which varied from $\sim$1.0--1.75$\arcsec$ over the 3.5-hour baseline. We attribute these systematics to flux contamination from nearby sources, since \name{} lies in a moderately crowded region of the Magellanic Clouds. In particular, a relatively bright (though redder) source lies $\sim$2$\arcsec$ to the south (Extended Data \cref{fig:image-cutouts}). We therefore adopted PSF photometry.

Our PSF photometry consisted of two steps. First, we used the \textsc{hipercam} pipeline to measure source centroids and local background levels in each frame for (i) \name{}, (ii) the two nearby sources to the south, and (iii) 10 reference stars. We used the \textsc{hipercam} pipeline for this step because it estimates the background in an annulus around each source (radii of 15--25 pixels) while masking pixels belonging to other known objects, which is essential in crowded Magellanic Cloud fields. The reference stars were required to be relatively isolated and detected in all three bands.

Second, using these centroids and background levels, we performed PSF photometry with the \texttt{phot}, \texttt{psf}, and \texttt{allstar} routines of \textsc{daophot} as implemented in IRAF. For each image, we built an independent PSF model by fitting a Moffat function to the 10 reference stars. We then fit \name\ and the two nearby sources simultaneously, integrating the \gls{PSF} out to a radius of 11 pixels (3.85$\arcsec$).

Rather than perform an absolute calibration of \name{} source photometry, we perform relative photometry using the same 10 reference stars that define the PSF model, which corrects for transparency variations caused by changing weather and seeing. 
We define a relative baseline for each star by identifying a 10-minute interval in the observations (40–50 minutes after the start) in which the reference-star photometry is particularly stable. The mean magnitude of each reference star in this interval was adopted as its baseline. For each image, we computed the offset between the baseline and PSF-measured magnitude of each reference star, and corrected the photometry of \name{} by the mean of these offsets.  The uncertainty on each epoch is taken as the quadrature sum of the statistical photometric error and the standard deviation of the reference-star offsets.

Across all three \Ultracam\ bands the source shows variability on the level of $\lesssim$2.5\% over a timescale of $\sim$2 hours, with the $g$-band light curve showing the least amount of scatter (Extended Data \cref{fig:uc-all}, lower three panels). This is not due to the sky variation between epochs as the mean FWHM in the top panel is nearly identical in $u$ and $g$, and is best in $r$ which does not show the least amount of scatter. The difference instead reflects signal-to-noise and systematics. The $u$-band has the lowest throughput and the strongest atmospheric extinction, so photon noise is higher and differential photometry against redder reference stars retains residual airmass dependent trends. In $r$, the source is much fainter relative to the sky background (Extended Data \cref{fig:image-cutouts}), which reduces the signal-to-noise.
The $g$-band balances detector efficiency, atmospheric extinction, and signal from a blue source, giving the most stable differential photometry. We therefore proceed with our analysis using the $g$-band light curve.

\paragraph{Detrending the light curve}
The unphased light curves shown in Extended Data \cref{fig:uc-all} show clear variability with two amplitudes overlapping the quadrature phases at 0.25 and 0.75 when phased by the orbital period. We attribute this variability to ellipsoidal modulation, in which tidal forces from a nearby companion distort the shape of the stripped star. We note that these amplitudes are also asymmetric, with a fractional flux compared to the median of roughly 0.4 and 1.8 for each amplitude respectively. One possible interpretation of this asymmetry could be Doppler beaming, which is expected in systems with similarly high orbital velocities \citep{Maxted2000}. However the fractional flux we observe is much higher than other systems exhibiting Doppler beaming, which is typically on the order of $\sim10^{-4}$ \citep[e.g.][]{Marten2010,Bloemen2011}. Furthermore, \name\ is much hotter than these systems, and the flux variation is expected to decrease with higher temperature \citep[e.g.][]{Bloemen2011,Bloemen2012,Breton2012,Herrero2014,Loeb2003,Marten2010,Zucker2007}.
The asymmetry therefore more likely reflects an observational systematic. In particular, the target was setting during the observations, and the color of the reference stars used to calibrate the photometry likely differs from that of the target. The two then suffer different atmospheric extinction, which changes with airmass, producing a one-way trend over the night.
We therefore proceed focusing only on the variation caused by ellipsoidal modulation.
To isolate this effect, we de-trend the light curve by fitting a linear model to the data and generating 30 realizations of the slope and intercept. These are obtained by randomly sampling from a multivariate normal distribution defined by the best-fit parameters and their covariance matrix with a median slope and intercept of 0.10 and 0.99 respectively. The de-trended $g$-band \Ultracam{} light curve is shown phase folded by the orbital period in \cref{fig:ellcjoker}.

\paragraph{Modeling the Ellipsoidal Variation} 

To model the ellipsoidal modulation in the $g$-band \Ultracam{} light curve, we use the \ellc{} light-curve modeling code \citep{Maxted2016}. We assume the stripped star rotates synchronously with the orbit, as expected for close binaries where tidal forces rapidly synchronize the rotation. The orbital period and mass function are taken from the circular spectroscopic solution.

As we have established that the companion must be a compact object, we model the companion star as a sphere with a fixed radius of $R_{\rm co}=10$ km, typical of a neutron star. Because $R_{\rm strip}\gg R_{\rm co}$, this choice does not affect the modulation amplitude, and the compact object contributes negligible surface brightness. The shape of the stripped star is set by the Roche potential, so it can deform in response to the companion. We assume that neither the companion nor any other source contributes additional flux.
For the stripped star we adopt a linear limb-darkening coefficient $\mu=0.2$ and a gravity-darkening coefficient $\tau=0.3$, following \citep{Gotberg2020B}, who applied the estimates of \citep{Claret2017} for hot stars to Galactic stripped stars in binaries with compact objects. 
We then generate 500,000 realizations of our system where we draw parameters from the following distributions: (i)~the stripped star mass, $M_{\rm strip}$, is drawn from a split-normal distribution based on the measured evolutionary mass described above, (ii)~the binary mass function $f_m$, from the spectroscopic posterior; and (iii)~the inclination $i$, from an isotropic prior (uniform in $\cos i$), truncated at $i>1\degree$ so that $M_{\rm co}$ remains finite. For each draw, we then compute the companion mass $M_{\rm co}$ from the mass function given $M_{\rm strip}$ and $i$. Finally, for each draw we assign the stripped star a radius, $R_{\rm strip}$, by adopting the effective radius, $R_{\rm eff}$, from the model in the posterior of the spectroscopic fits to \name{} performed by \cite{Gotberg2023} that has the closest evolutionary mass to the adopted $M_{\rm strip}$. We adopt this approach, instead of drawing from a separate distribution for $R_{\rm strip}$, because the two properties are highly correlated in  \cite{Gotberg2023}.

For each realization of our system, we then compute an \ellc{} model and evaluate $\chi^2$ against one of the 30 detrended light curves described above, chosen at random. Data normalization uncertainty is therefore absorbed into our analysis. To generate probability distributions for the input parameters, we then resample 100,000 models from the 500,000 draws and assign each  weights based on their likelihood: $\mathscr{L}=\exp\bigl[-0.5\,(\chi^2-\chi^2_{\rm min})\bigr]$. The joint $M_{\rm strip}$--$i$ density and the companion-mass CDF in \cref{fig:kde} are computed from these weighted samples. 
Models that reproduced the observed amplitude of the ellipsoidal modulations show clear correlation between $M_{\rm strip}$ and inclination angle, while inclinations below $16.9$\degree\ and $11.4$\degree\ lie outside the 68\% (gold) and 95\% (cyan) Highest Posterior Density (HPD) regions, respectively. The resulting companion mass distribution has a  marginal posterior that peaks at \MCompanionKDE~\Msun\ with a 68\%\ HDI of [\MCompanionHDIse]~\Msun\ and a 95\%\ HDI of [\MCompanionHDInf]~\Msun.

\paragraph{X-ray Upper Limits}

The field of \name{} was observed with the \Chandra\ X-ray Observatory Advanced CCD Imaging Spectrometer (ACIS-S) in VERY FAINT mode across two epochs: 2025 August 22 (ObsID 30583; 27.70 ks) and 2025 October 4 (ObsID 28037; 18.77 ks), yielding a combined exposure of 46.47 ks. 
Each event file is reprocessed against the latest calibration database using \emph{chandra\_repro} within version 4.18 of the Chandra Interactive Analysis of Observations (CIAO) software \citep{Fruscione2006,fruscione_ciao_2026} to produce calibrated level-2 event files. Aperture photometry is performed with the CIAO \emph{srcflux} tool at the optical position of the source, using a 3\arcsec-radius circular source aperture (6.10 pixels at the ACIS-S plate scale of 0.492\arcsec\,pixel$^{-1}$) and a concentric 10\arcsec–20\arcsec\, background annulus centered on the same coordinates. For each observation, \emph{srcflux} returns the source and background counts, aperture areas, exposure time, and the PSF encircled-energy fraction (via \emph{psfmethod=arfcorr}). We adopt the standard hard band (2.0–7.0 keV) used by the Chandra Source Catalog \citep{Evans2010,Evans2024}, as this range balances integrated effective area against background, which rises rapidly above this energy on ACIS-S.

No X-ray emission is detected at the optical position of \name{}. We therefore used \emph{aplimits} to derive a detection upper limit following \citep{Kashyap2010}. We adopt the one-sided 3-$\sigma$ Gaussian equivalent (99.87\%) given that true flux values are non-negative. For \emph{aplimits}, this means we set a  3$\sigma$ probability upper-limit on the count rate ($\alpha = 0.0027$) for a source to be missed 50\% of the time ($\beta = 0.5$). We corrected the resulting aperture rate limit for the enclosed PSF fraction in each observation. 
Assuming an absorbed power-law spectral model with photon index $\Gamma = 2.0$ and Milky Way neutral hydrogen column density of NH$_{\rm{MW}} = 3.62\times 10^{21}$\,cm$^{-2}$ \citep[][]{HI4PICollaboration2016}, we estimate an unabsorbed flux upper limit of $\mathrm{F}_{2-7\mathrm{keV}}$\ChandraFluxUnabsorbed\,\ergscm.
Given the distance to the SMC \citep[60.6 kpc;][]{Hilditch2005} 
and $L= 4\pi d^2 F$, the corresponding luminosity is $\mathrm{L}_{2-7\mathrm{keV}}$\ChandraLumUnAbsorbed\,\ergs. 
We include our estimated X-ray upper limits with and without absorption in Table~\ref{tab:params}.

Additionally, we searched archival XMM-Newton observations for X-ray emission from \name. Fourteen observations obtained between 2001 and 2016 cover its position, with a combined exposure of $\sim$308\,ks after filtering for periods of high background. We reduced them with the \XMM\ Science Analysis System (SAS) version 20.0.0, using only the EPIC-pn detector (medium filter) for its high sensitivity and large effective area. We selected single and double events (PATTERN $\leq$ 4), applied the standard
\#XMMEA\_EP flag screening, and removed flaring intervals using count rate histograms.
We extracted counts with \texttt{funcnts} in \textsc{FunTools} \citep{Mandel2011} using a 30\arcsec-radius aperture at the optical position of \name\ with a 90\arcsec-radius source-free aperture for the background. We derived a one-sided 3$\sigma$ upper limit with \emph{aplimits}, as for the \Chandra\ upper limit, correcting for the enclosed fraction of the source ($85\%$). \name\ is not significantly detected in any individual observation or in the stack of all 14, which we made with the SAS task \textsc{merge}. Assuming the same $\Gamma$, NH$_{\rm{MW}}$ and 2--7\,keV band, the stacked $3\sigma$ unabsorbed flux upper limit is \XMMFluxUnabsorbed\,\ergscm, corresponding to a luminosity \XMMLum\,\ergs. The XMM-Newton limit is comparable to our Chandra limit, and slightly
lower, because the longer exposure and larger collecting area of XMM-Newton
outweigh the lower background of Chandra. We therefore adopt the Chandra upper limit, which is more conservative (i.e., less stringent) than the XMM-Newton limit and is less susceptible to source confusion because of Chandra’s narrower PSF.

\paragraph{X-ray Modeling}
Under the standard \gls{BHL} wind-accretion regime \citep{BondiHoyle1944}, the X-ray luminosity assuming a \gls{NS} companion can be approximated by: 

\begin{equation}
L_{\mathrm{BHL}}
=
\epsilon \frac{G \, M_{\mathrm{NS}} \, \dot{M}_{\rm acc}}{ R_{\rm NS}}
\end{equation}

\noindent where $M_{\rm NS}$ and $R_{\rm NS}$ are the neutron star mass and radius, for which we adopt 1.4\,\Msun\, and 10\,km respectively, along with a radiative accretion efficiency of $\epsilon \sim \mathrm{G}M_{\rm NS}/R_{\rm NS} c^2 \sim 0.2$. The accretion rate is given as

\begin{equation}
\dot{M}_{\rm acc}
=
\frac{\alpha}{2}
\frac{(G M_{\mathrm{NS}})^2 \, \dot{M}_{\mathrm{wind}}}
{v_{\mathrm{rel}}^{3} \, a^{2} \, v_{\mathrm{wind}}}
\end{equation}

\noindent for which we adopt the upper limit mass-loss rate of $\dot{M}_{\mathrm{wind}}=10^{-9}\;\mathrm{M}_\odot/\mathrm{yr}$ from \citep{Gotberg2023}, and compute the orbital separation $a$. The relative velocity ($v_{\rm rel}$) between the neutron star and stellar wind is the combined wind ($v_{\rm wind}$) and orbital ($v_{\rm orbit}$) velocities added in quadrature. The wind velocity is calculated with a standard $\beta$-law ($v_{\rm wind}=v_{\infty}(1-\frac{R_{\rm *}}{a})^\beta$) where we assume $\beta=1$ and the terminal velocity to be $v_{\infty}=1.5\,v_{\mathrm{esc}}$ \citep{Lamers1995}. Finally, we assume the Bondi--Hoyle correction factor $\alpha$=1.5.

In order to compare our bolometric X-ray predictions to our band specific upper limit, we also need to consider what fraction of X-ray light would fall into the 2--7 keV \Chandra\ band. When calculating the upper limit, we assumed the X-ray SED would follow a power-law distribution with a photon index of $\Gamma=2$ which allowed us to convert photon counts to flux. For NS X-ray binaries the transition between systems that follow a power-law versus those that have an additional thermal component generated by the disk is typically taken when the X-ray luminosity exceeds $5\%$ of the Eddington luminosity which is given as $L_{edd}=4\pi c\mathrm{G}M_{\mathrm{NS}}/\kappa$ where $\kappa=0.2(1+X_H)\mathrm{cm}^2\mathrm{g}^{-1}$. Adopting $X_H=0.1$ from the spectroscopic fit \citep{Gotberg2023}, gives an Eddington Luminosity of $L_{\mathrm{X,Edd}} \sim 3.2\times10^{38}\,\mathrm{erg\,s^{-1}}$, which is five orders of magnitude above our upper-limit thereby justifying the power-law model. However, given that there are no X-ray detections for intermediate mass stripped stars the exact slope of the power law and therefore the fraction of X-ray light captured by the 2--7 keV band is unconstrained. We therefore adopt averaged empirical values from low luminosity X-ray NS binaries which includes the slope $\Gamma=2$  \citep{Wijnands2015}, and the bolometric-to-band conversion factor $\eta=0.21$  \citep{Fragos2013}.

Folding in the spectroscopic and orbital parameters of \name{} along with these assumptions results in a bolometric X-ray luminosity of $L_{\mathrm{X,BHL}} \sim 1.5\times10^{33}\,\mathrm{erg\,s^{-1}}$ which is just above our unabsorbed $3\sigma$ \Chandra\  upper limit of $\mathrm{L}_{2-7\mathrm{keV}}$\ChandraLumUnAbsorbed\,\ergs, suggesting the source should have been marginally detectable. 
However, this estimate is highly sensitive to the wind properties of the stripped star which are difficult to constrain given the lack of emission features typically used to estimate the wind mass-loss rate in optical spectra. For instance, a mass-loss rate of $0.49\times10^{-9}\,\mathrm{M_\odot\,yr^{-1}}$ is sufficient to push the predicted emission below detectability.

Theoretical limitations of the wind properties have a strong impact as well. If we instead adopt $v_\infty = 2.5\,v_{\mathrm{esc}}$, as used by \citep{Gotberg2023} following \citep{Lamers1995}, the predicted X-ray luminosity is instead  $\sim2.4\times10^{32}\,\mathrm{erg\,s^{-1}}$ which is  below our fiducial value but similar to the X-ray luminosity for the Galactic hot subdwarf X-ray binary HD-49798 \citep{mereghettiDiscoverySpinupXray2016}. The accretion efficiency is also highly uncertain: at these low inferred luminosities, accretion may instead proceed through an advection-dominated flow, which could substantially reduce the X-ray luminosity relative to the standard assumption of $\epsilon = 0.2$ \citep{Qiao2021}.

Compounding the difficulty of predicting X-ray luminosities from systems such as \name{}, the \gls{BHL} formalism neglects both the spin and magnetic field of the NS. We expect that the characteristic lifetime for an NS companion would be well below the total lifetime for the currently present intermediate mass stripped star (predicted to be order $
\lesssim100$ Myr \citep{HovisAfflerbach2025}) and would therefore likely have non-negligible properties in this regard. This may indeed invoke the ``propeller regime'' \citep{Illarionov1975} in which the magnetosphere of a rapidly rotating neutron star prevents infalling material from reaching the surface and therefore quenches X-ray emission.  

\paragraph{Binary evolution modeling -- Model set-up}
We use the binary stellar evolution code MESA \citep{Paxton2011, Paxton2013, Paxton2015, Paxton2019, Jermyn2023}, version 24.08.1. All our inlists and run\_star\_extras files are made publicly available. We adopt the following assumptions and settings in our models. If not mentioned below, we adopt MESA's default settings. We use the Ledoux criterion to determine convection and set the semiconvective mixing \citep{Langer1985} efficiency to $\alpha_{\rm sc} = 1$. We adopt the mixing length theory \citep{BohmVitense1958} and set the mixing length to $\alpha_{\rm MLT} = 1.75$ times the pressure scale height, motivated by observations of red giant branch stars in clusters \citep{Joyce2018, Joyce2023}. We further implement convective overshoot above convective regions by extending them with 0.275 times the pressure scale height \citep{Anders2022, Johnston2019}. 

We choose to use the nuclear network \texttt{approx21.net}, which is created for computations until core collapse but with simplified nucleosynthesis \citep{Timmes1999, Paxton2011}. When running a test with the larger nuclear network \texttt{mesa\_49.net}, we see small differences related to the location of convective layers during the late evolutionary stages. Because the core masses are not affected, we choose to use the smaller \texttt{approx21.net} network, which results in faster computations that converge more easily. For the atmosphere boundary, we adopt the default Eddington $T-\tau$ relation.

We implement wind mass loss during the main sequence evolution by adopting the theoretical scheme from \cite{Vink2001} and switch to the more generic and empirically determined scheme from \cite{DeJager1988} when the surface temperature decreases below 10 kK. However, we switch to the helium star wind scheme from \cite{Vink2017} if the surface temperature is higher than 10 kK and surface hydrogen mass fraction less than 0.4. 
This weak helium star wind matches better the constraints from observations \citep{Gotberg2023}, compared to the extension of the Wolf-Rayet scheme by \cite{Nugis2000} as is the default implementation in MESA's ``Dutch'' wind scheme. 

More information regarding model design is found in the Supplementary Information.

\subsection*{Data availability}\label{sec:data}
The \textit{Chandra} and \textit{XMM-Newton} observations analysed in this
study are publicly available from the Chandra Data Archive
(\url{https://cxc.cfa.harvard.edu/cda/}) under observation IDs 28037 and 30583 and the XMM-Newton Science Archive
(\url{https://nxsa.esac.esa.int/}) under observation IDs 0018540101, 0110000201, 0304250401, 0304250501, 0304250601, 0601210501, 0601210601, 0780090301, 0780090401, 0780090501, 0780090601, 0780090701, 0780090801 and 0780090901. Additional derived data products that support the findings of this study, including optical spectroscopy and images, will be deposited in a public repository and made available upon acceptance of the manuscript.

\subsubsection*{Code availability}\label{sec:code}
The code used to perform the analysis and create the figures in this project will be released publicly upon publication. 
Orbital analysis was performed with \texttt{The Joker} 1.3.1 \citep{Joker12017,Joker22017} in combination with \texttt{Python} 3.11.15, \texttt{PyMC} 5.28.5 \citep{PyMC}, \texttt{PyTensor} 2.38.3, \texttt{pymc-ext} 1.0.1, \texttt{ArviZ} 0.23.4 \citep{Arviz}, \texttt{twobody} 0.9.1, \texttt{NumPy} 2.4.6 \citep{harris2020array}, \texttt{SciPy} 1.17.1 \citep{2020SciPy-NMeth}, \texttt{Astropy} 8.0.1 \citep{astropy:2013,astropy:2018,astropy:2022}, \texttt{pandas} 3.0.5 \citep{mckinney-proc-scipy-2010,reback2020pandas}, \texttt{Matplotlib} 3.11.1 \citep{Hunter:2007}, \texttt{corner} 2.3.0 \citep{Corner} and \texttt{dill} 0.4.1. 
\Ultracam\ \citep{Dhillon2007} data were reduced with \texttt{hipercam}.
Light curves were modelled with \texttt{ellc} 1.8.5 \citep{Maxted2016} in combination with \texttt{Python} 3.11.14, \texttt{NumPy} 1.26.4, \texttt{SciPy} 1.16.3, \texttt{Astropy} 7.2.0, \texttt{pandas} 2.3.3, \texttt{Matplotlib} 3.10.8 and \texttt{ipywidgets} 8.1.8. 
XMM-Newton data was reduced with XSPEC 12.9.1 \citep{Xspec}. 
Chandra data were reduced with \texttt{CIAO} 4.18.0 \citep{Fruscione2006,fruscione_ciao_2026}, \texttt{Python} 3.12.12 and \texttt{Sherpa} 4.18.0 \citep{Sherpa}. 
The binary evolution models were computed with \texttt{MESA} 24.08.1 \citep{Paxton2011,Paxton2013,Paxton2015,Paxton2018,Paxton2019,Jermyn2023}. 
Spectral disentangling was performed with the shift-and-add code of Shenar, Fabry \& Bodensteiner, version 2.0 (11 September 2023), following \citep{Gonzalez2006,Shenar2020LB1,Shenar2022}. This disentangling along with the remaining analysis were run with \texttt{Python} 3.13.13, \texttt{NumPy} 2.4.4, \texttt{SciPy} 1.17.1, \texttt{Astropy} 7.2.0, \texttt{pandas} 2.3.3, \texttt{Matplotlib} 3.10.9, \texttt{specutils} 2.3.0 \citep{specutils}, \texttt{cmasher} 1.9.2 \citep{Cmasher}, \texttt{astroquery} 0.4.11 \citep{astroquery:2019} and \texttt{ipywidgets} 8.1.9.

\section{Acknowledgements}
We thank the team at Las Campanas Observatory and the Observatories of the Carnegie Institution for Science, which resulted in our multi-year binary orbital monitoring program of \name.
We thank Christopher D. Matzner, Joshua S. Speagle, Marten van Kerkwijk, Tomer Shenar, Gwen Rudie, Peter Senchyna, Lisa Blomberg, Kareem El-Badry, Jakub Klencki, Abigail Denney, Dandan Wei, Debasish Dutta, Koushik Sen, Kunal Deshmukh, and Javier Mor\'{a}n-Fraile for helpful conversations. 
We thank the Munich Institute for Astro-, Particle and BioPhysics (MIAPbP) for the 2022 program ``The Fundamental Role of Stellar Multiplicity in Stellar Dynamics and Evolution'', where discussions contributed to the development of this work. 

B.L. and H.S. have  received financial support from the Flemish Government under the long-term structural Methusalem funding program by means of the project SOUL: Stellar evolution in full glory, grant METH/24/012 at KU Leuven.
M.R.D. acknowledges support from NSERC through grant RGPIN-2025-06224, the Canada Research Chairs Program (CRC-2023-00127), the Ontario ERA program (ER22-17-164) and the Dunlap Institute at the University of Toronto.
A.L. acknowledges support from the NSERC and is funded through a NSERC Canada Graduate Scholarship—Doctoral. A.L. is also supported by the Data Sciences Institute at the University of Toronto through grant number DSI- DSFY3R1P02. 
K.A. acknowledges support from the Australian Research Council Centre of Excellence for Gravitational Wave Discovery (OzGrav), through project number CE230100016. 
A.V.-G. is supported by the Carlsberg Foundation, grant CF24-1046. V.S.D. is supported by a Leverhulme Research Fellowship (RF-2025-297/9), the Spanish Ministry of Science, Innovation and Universities (PID2023-151588NB-I00), and the Agencia Estatal de Investigaci{\'o}n (AEI) through the Severo Ochoa Centre of Excellence accreditation awarded to the Instituto de Astrofísica de Canarias, grant CEX2025-001609-S, funded by MICIU/AEI/10.13039/501100011033. 
K.B. acknowledges support from the Falco DeBenedetti Early Career Professorship.
E.L. acknowledges funding through a start-up grant from the Internal Funds KU Leuven (STG/24/073), a Veni grant  (VI.Veni.232.205) from the Netherlands Organization for  Scientific Research (NWO), and the Research Foundation – Flanders (FWO) under the Odysseus Program, Type II (G0AT525N).
Support for this work was provided by the National Aeronautics and Space Administration through Chandra Award Numbers GO2-23031A and GO3-24037X issued by the Chandra X-ray Center, which is operated by the Smithsonian Astrophysical Observatory for and on behalf of the National Aeronautics Space Administration under contract NAS8-03060.

\section*{Author contributions}
B.L. wrote the manuscript with contributions from M.R.D., Y.G., K.A., and A.V.G and comments from all authors.
The MagE spectra were obtained by B.L., A.L., M.R.D. and Y.G. M.R.D. reduced the spectra. A.L. measured the radial velocities and stacked the spectra. B.L. performed the orbital analysis and spectral disentangling.
The \Ultracam\ observations were proposed by Y.G. and C.J. and obtained by Y.G., C.J. and V.D. M.R.D. reduced the data and performed the photometry. B.L. performed the light-curve analysis.
K.A. computed the \XMM\ upper limit. \Chandra\ observations were proposed by B.L. who reduced the data and computed the \Chandra\ upper limit and the X-ray models.
Y.G.\ and A.V.G. designed, computed  and analyzed the MESA evolutionary models, and also integrated the future orbital evolution of the models to double neutron star merger. M.R.D. estimated companion rates from POSYDON models.

\section*{Supplementary Information}

\subsection{Updated stacked spectrum}

After using the best-fit orbital solution (Table~\ref{tab:params}, \cref{fig:ellcjoker}) to correct for the radial velocity shifts, we interpolate each epoch spectrum onto a new wavelength grid and stack the spectra. The method is described in detail in \cite{Laroche2026}. We present the resulting optical spectrum stack in Extended Data \cref{fig:spectrum}. 
The figure shows that \name{} clearly exhibits the following spectral lines: \HeII{} 6560 + H$\alpha$, \HeII{} 5411, N\textsc{v} 4945, \HeII{} 4860 + H$\beta$, \HeII{} 4686, \HeII{} 4542, \HeII{} 4339 + H$\gamma$, \HeII{} 4200, \HeII{} 4100 + H$\delta$, \HeII{} 4026, along with the two interstellar features Ca\textsc{ii} H \& K.  
This is in line with the description in \cite{Gotberg2023}, but the signal-to-noise ratio per pixel in this updated stack is $\sim 60$ and therefore somewhat better than the previous $\sim 50$.

\subsection{Rate Estimates for Double Stripped Stars} 

The existence of double stripped stars has been considered in the literature before. \cite{Pols1994} described how the supernova order in binary systems can be reversed by conservative mass transfer. Because the system was sufficiently tight to interact to begin with, this suggests that two stripped stars must be formed and co-exist before the initially less massive star can explode prior to the initially most massive star. Similar evolutionary pathways led to the small percentage of stripped companions to stripped supernovae that \cite{Renzo2019} and \cite{Souropanis2026_powerofbinaries,Zapartas2026} found in their population synthesis simulations. 
In particular, an analysis of the solar-metallicity POSYDON population underlying these works \citep{Souropanis2026_powerofbinaries,Zapartas2026} shows that, among Type Ibc supernova progenitors with final stripped star masses below 8 M$_\odot$ (15,845 systems; Type IIb progenitors excluded), only about 0.6\% (92 systems) have a stripped star companion at the time of explosion. These systems arise from roughly equal mass binaries, spanning a wide range of initial masses, and the lowest-mass stripped star companion in this population is 2.5 M$_\odot$. White dwarf companions are rarer still, at about 0.09\% (15 systems). These all descend from near equal mass binaries with initial primary masses of 6--7 M$_\odot$ and secondaries of 5--7 M$_\odot$, and they are massive ($>1$ M$_\odot$) and paired with low-mass progenitors (final stripped star masses of 1.7--2.5 M$_\odot$). In contrast, about 10\% of the population (1619 systems) have a NS or BH companion. Within this population, a compact object companion is therefore more than an order of magnitude more likely than a stripped star companion, and white dwarf companions to stripped stars that are massive enough to explode are very rare. These fractions come from a single population synthesis code with one set of binary physics assumptions (notably relatively non-conservative mass transfer) at solar rather than SMC metallicity, so we treat them as order of magnitude guides rather than precise rates. We further note that different population synthesis studies using other codes report similar rates, suggesting a general consensus at the order of magnitude scale \citep{Ercolino2026, Martinez2025}.

\subsection{Modeling the future evolution of \name{}}

We forecast the evolution of the system by computing a binary evolution model. The detailed setup is described in \method{}, while here we go through the details regarding the model design and evolution. As we show below, it is likely that the system will form a double neutron star that will merge via gravitational-wave emission within the age of the Universe.

\paragraph{Creating a stripped star model}
To model the future evolution of the binary, we first create a stripped star model that matches the observationally constrained properties of the current system: $T_{\star} = 72^{+76}_{-8}$ kK, $\log_{10} (g_\star/\mathrm{cm\,s}^{-2}) = 5.4^{+0.6}_{-0.5}$, $X_{\rm H, \, surface} = 0.10^{+0.25}_{-0.09}$, $X_{\rm He, \, surface} = 0.90^{+0.09}_{-0.25}$, $\log_{10} (L/L_\odot) = 4.15^{+0.85}_{-0.17}$ \cite{Gotberg2023}, where the values marked $\star$ are expected to more closely correspond to MESA's photospheric values compared to the effective values published from the spectroscopic analysis \citep[cf.][]{Groh2008, Groh2014}. To do this, we compute a zero-age main sequence model with 12\,\Msun\ and Z = 0.002 to match the SMC \citep{Brott2011A}, adopting solar-scaled abundances \citep{Grevesse1998}. We place the model to orbit an 8\,\Msun\ point-mass companion every 20 days. This leads to mass transfer initiated early during the crossing of the Hertzsprung gap. To resolve this mass-transfer episode, we use the implicit scheme of Kolb \citep{Kolb1990}, assuming that the mass transfer is conservative throughout the episode. We save a model when helium core burning has started and the central helium mass fraction ($X_{\mathrm{He, \, center}}$) has decreased below 0.9. At this stage, the binary has detached and the stripped star is contracting.

The binary system described in this study has a very short orbital period and is most likely the outcome from common envelope evolution. Although it is possible to model common-envelope evolution in MESA \citep[e.g.,][]{Clayton2017, Fragos2019, Marchant2021, GallegosGarcia2023, Bronner2024}, the treatment remains computationally challenging and of limited accuracy because the underlying physics is fundamentally three-dimensional. We, therefore, opt for adjusting the model stripped through mass transfer that we described above until it matches the observed properties.

The model that we saved after mass transfer has 4.25\,\Msun\ and if continued to halfway through central helium burning (defined as when the central helium mass fraction is $X_{\rm He, center} = 0.5$), it has an effective temperature of 61 kK, $\log_{10} (g/\mathrm{cm \, s}^{-2}) = 4.7$ and a surface hydrogen and helium mass fraction of 0.39 and 0.61 respectively. This is too cool, too fluffy, too hydrogen-rich and too helium-poor compared to the observations, which likely means that the observed star has been stripped deeper than the model. We therefore artificially remove material from the surface of the model using the control parameter \texttt{mass\_change} set to $-0.0001$. Since the observed surface hydrogen and helium mass fractions are somewhat uncertain, we halt the mass removal at three stages: when the surface hydrogen mass fraction has reached 0.3, 0.2, and 0.1. Evolving each of these to halfway through the central helium burning phase, we find their effective temperatures and surface gravities to reach 71, 81, 87 kK and $\log_{10} (g/\mathrm{cm \, s}^{-2}) =$ 5.1, 5.4, 5.5 respectively. Their surface abundances do not change much during the evolution because of the low wind mass loss adopted for the stripped stars, which amounts to $\sim 10^{-8}$ M$_\odot$/yr during the stripped phase \citep{Vink2017} and that is motivated by the observed absorption spectrum \citep{Gotberg2023}. In the spectral fitting of the star \citep{Gotberg2023}, the effective temperature was the least constrained parameter, because of the lack of He~\textsc{i} lines to compare with the He~\textsc{ii} lines, while both the surface gravity and surface hydrogen mass fraction had smoother and narrower $\chi^2$ distributions with clearly peaked minima. We therefore consider the model with $X_{\rm H, \, surface} = 0.2$ and $\log_{10} (g/\mathrm{cm \, s}^{-2}) = 5.4$ to be the best fit. This model's surface hydrogen mass fraction is 0.1 higher than the best-fit, but still well within the errors of the measurement. It has a luminosity of $\log_{10} (L/L_\odot) \sim 4.2$ and a mass of 3.6 M$_\odot$. This is very close to the best-fit evolutionary mass that we presented for the star in \citep[$M_{\rm evol} = 3.38^{+4.34}_{-0.42}$ M$_\odot$,][]{Gotberg2023} and that we use in the main text of this study, making the model suitable for exploring the future evolution of the system. We show its evolution in the Hertzsprung-Russell diagram and Kiel diagram in Extended Data \cref{fig:stripped_model}, which we describe in more detail below. 

It is likely that a model with a slightly different mass can also be considered well-fitting, since the observationally constrained parameters have uncertainties. The model parameters also change, albeit little, as the star evolves, and we do not know how far the star has reached through its central helium burning. However, given its reasonable match, we continue our study using this model and evolve it in a binary with 4.17 hours until central oxygen depletion, defined as when the central oxygen mass fraction ($X_{\rm O, \, center}$) decreases below 0.01\footnote{The exceptions are models D and G, which stop at $X_{\rm O,\, center} = 0.0108$ and 0.071, respectively, see Table~\ref{tab:model_variations_a}.}. Oxygen burning takes only about 4 years and the subsequent burning phases are even shorter. This means that when we stop the evolution, the model is so close to core collapse that the star's surface parameters and envelope structure do not have time to evolve further and the outer structure is representative of the structure at core collapse. Compared to progressing the models to core collapse, this stopping condition reduces the computational resources. We do not account for the effect of gravitational-wave radiation on the orbital evolution while the star is still alive, because, for this orbital configuration and at this evolutionary stage, the effect is minimal and should only change the orbit by $\sim 1$ second. 

In our simulations, we adopt high spatial and temporal evolution by setting \texttt{mesh\_delta\_coeff = 0.6} and \texttt{varcontrol\_target = 5d-4}. We also include a set of additional resolution parameters because we find smoother numerical evolution.

\paragraph{Modeling case BB mass transfer}
In all our models, the stripped star fills its Roche lobe when it expands during helium shell burning because the orbital separation is so small, that is, the models experience case BB type mass transfer \citep[e.g.,][]{Dewi2003,VignaGomez2018}. For this reason, the orbital evolution is directly dependent on what mass is assumed for the companion, and how angular momentum loss is treated during mass transfer. We found that the companion has to be a compact object that most likely has a mass of $1.4$\,\Msun\ (see main text), which is consistent with the mean of the observed mass distribution of Galactic double neutron stars \citep{Farrow2019}. We therefore adopt this as the companion mass in our fiducial model, which we refer to as model A. We also compute a model with a companion mass of 3 M$_\odot$ (model B). We change the angular momentum loss treatment in model C to a circumbinary disk, and in model D to material leaving the system with angular momentum from the center of mass of the donor (Jeans loss) \citep{Tauris2023}. Model E adopts a pure helium star model to represent the stripped star, for making a better comparison with what most binary population synthesis codes would predict. Models F and G use more massive stripped star models of 4.28\,\Msun{} and 4.84\,\Msun{}, which we created from initially 13.5\,\Msun{} and 15\,\Msun{} stars. The stripped stars in models F and G are created in the same way as described above and stripped down to a surface hydrogen mass fraction of 0.2. We list the model variations in Table~\ref{tab:model_variations_a}.

Compact objects usually do not accrete substantial amounts of material when fed through Roche lobe overflow by a more massive donor, and just $\sim10^{-3}$\,\Msun\ is enough to spin up the neutron star to milliseconds \citep{Tauris2012, Tauris2015, Tauris2017}. We limit the accretion to the Eddington accretion rate in the models that have a neutron star accretor and that use isotropic re-emission\footnote{This is simply because the function already was implemented in MESA for that setup.}(models A, E, F, and G). In our models the Eddington accretion rate reaches $\sim 10^{-8}$\,\msunyr. In Extended Data Fig.~\ref{fig:Mdot_a}, we show the mass change rate from the stripped star during the case BB mass transfer as function of time and for models A, B, C and D. As the figure shows, the average mass transfer rate is $\sim 10^{-5} - 10^{-4}$\,\msunyr, which means that the mass transfer is highly non-conservative. Because of this inefficient accretion, material leaves the system during the case BB mass transfer, carrying away angular momentum. The amount of mass and angular momentum removed from the system sensitively impacts the evolution of the orbit. Non-conservative mass transfer is not very well understood and has very few observational constraints, largely because of the scarcity of observed post-mass transfer systems \citep{WeiDandan2026,Laroche2026}. 

In MESA, angular momentum loss during non-conservative mass transfer is parametrized by the fraction of specific angular momentum ($j$) that is lost, that is $dj/dt = \gamma \, j$, where $\gamma$ can be either a constant or a quantity that changes during the interaction.  
The most standard assumption is that the material leaves from the system with the specific angular momentum of the center of the accretor star (so-called isotropic re-emission, \cite{vandenHuevel1973, soberman_stability_1997}), which means that $\gamma = M_{\rm donor}/M_{\rm accretor}$ and could, for example, be motivated as a jet carrying away the material. This is the method we adopt for our fiducial model and the mass variations (models A, B, E, F, and G). 
However, recent hydrodynamical simulations show that material tends to leave the system through the second Lagrange point (L2), carrying away more angular momentum compared to isotropic re-emission with an average $\gamma \sim 6-7$ \citep{Sherback2025_rapidmt, Sherbak2025}. Circumbinary disks are also considered to be efficient at draining angular momentum, with $\gamma = (M_{\rm total}^2/(M_{\rm donor}M_{\rm accretor})) \sqrt{a_{\rm ring}/a}$. They could for example be created through L2 outflows or from material leftover from the previous envelope-stripping, therefore perhaps providing the most similar mass transfer case to the hydrodynamical simulations. Above, we used $M_{\rm donor}$ and $M_{\rm accretor}$ to refer to the masses of the donor and accretor stars, $M_{\rm total} = M_{\rm donor} + M_{\rm accretor}$, $a$ is the orbital separation and $a_{\rm ring}$ is the separation between the binary's center of mass and the inner rim of the circumbinary disk. We adopt a value $a_{\rm ring}/a = 2$ in our circumbinary disk variation in model C. 

We find two different outcomes of the Roche lobe overflow in our models: either mass transfer continues at a rate that can be computed and the model reaches oxygen depletion during the case BB mass transfer, or the mass transfer rate becomes excessively large and we interpret this as the development of a common envelope. 
These different scenarios are clearly visible in the bottom panel of Extended Data \cref{fig:Mdot_a}, which shows the orbital separation plotted as function of time for the models A, B, C, and D, which include the different companion mass (B) and the different angular momentum loss scheme (C, D). The isotropic re-emission that we implement in our standard model (A) causes orbital tightening, while the Jeans loss causes widening (D), which is also the case for the isotropic re-emission when a more massive companion is included (B). In the model with a circumbinary disk, the orbit tightens and mass transfer rate increases. But eventually it tightens so quickly ($\dot{a} = -10^{-2.2}$\,\rsunyr) that the mass transfer increases to very high rates ($\sim10^{-2.5}$\,\msunyr), and we consider that the model will enter a common envelope. 

Here, we consider that a common envelope would lead to a merger because of the originally very tight orbit. Such mergers would most likely be associated with an electromagnetic transient, potentially a luminous fast blue optical transient \citep{Klencki2026, Neights2026}. But, if the companion is a neutron star, the transient could be faint and perhaps less easy to discover. Potential products of such mergers include a Thorne-\.Zytkow object with a helium envelope \citep{Nathaniel2025}, or complete disruption of the stripped star. The only model that leads to a merger during Case BB interaction is the circumbinary disk model (C). 

The other models continue the case BB mass transfer until the end of the simulations. Their orbits adjust during the mass transfer, leading to either tighter (models A, E, F, G) or wider (models B, D) orbits of maximum six hours. During mass transfer, the stripped star loses about a solar mass of material, but only a fraction is accreted by the companion. In our standard model (A), $0.00072$\,\Msun\ of the transferred material is accreted by the neutron star companion, which reaches a mass of $1.40072$\,\Msun. The accretion efficiencies are similar for the other models where we adopt Eddington-limited accretion (A, E, F, G).
The neutron stars thus grow minimally in mass. This is because the Eddington-limited accretion rate is so low  ($\sim 10^{-8}$\,\msunyr) in combination with the relatively short duration of the case BB mass transfer ($\sim 10^4$ years). We use the amount of accreted material to estimate the neutron star spin using the relation from \cite[][see their Eq.~14]{Tauris2012}, and assuming it was not spinning before mass transfer started. We obtain spin periods of order 100 milliseconds, which does not really classify as recycled, but rather regular spin periods of pulsars \citep{Tauris2023}. This frequency is also quite commonly found in double neutron star systems \citep{Tauris2023}.

Because the mass transfer is almost fully non-conservative, most of the transferred material is ejected to the surroundings, which in our models correspond to $\sim 1.2-1.4$\,\Msun{} and the vast majority of which is helium. Since it remains unclear how material is ejected from the system, we take a simplistic approach when estimating how this mass loss creates and shapes the circumstellar structure. We assume that the material is distributed spherically and ejected with the orbital speed of the neutron star. This results in a circumstellar density as displayed in Extended Data Fig.~\ref{fig:csm_density}. The figure shows that the density reaches $10^{-12}$g/cm$^{3}$ at the orbital separation ($\sim 10^{11}$cm) and drops relatively monotonically to $\sim10^{-25}$g/cm$^3$ at 1pc ($10^{18.5}$cm). These densities correspond to number densities in the range $n_{\rm He} = 10^{11}$-$0.01$ cm$^{-3}$.

\paragraph{The supernova and neutron star formation}

For all our models, all hydrogen is lost during Case BB mass transfer. But, at the end of the simulations, these models still have a non-negligible amount of helium left ($\sim 0.2 - 0.5 \, M_\odot$), which is consistent with models for hydrogen-free and helium-rich supernovae \citep{Dessart2012, Hachinger2012}. Given the on-going non-conservative mass transfer at core collapse and the creation of a dense circumstellar structure, we expect that the supernova will show signs of the circumstellar material, potentially appearing as a type Ibn \citep{Sollerman2020, Dessart2022}. We also note that the star is not spherical at explosion, which could alter its appearance and the supernova further \citep{DuPont2024}.

\cite{Tauris2015} used evolutionary models to find that helium stars with a metal core mass $\gtrsim 1.45 M_\odot$ reach iron core-collapse. For their pure helium stars, this limit corresponds to helium star masses $\gtrsim 2.6 M_\odot$. Similarly, \cite{Jiang2021} computed the evolution of a 3.2 M$_\odot$ pure helium star orbiting a 1.35 M$_\odot$ neutron star every 24 hours. They found the star to get stripped further during Case BB mass transfer and it produced a carbon-oxygen core with mass of 1.60 M$_\odot$. The model eventually reached core collapse and resulted in a 1.44 M$_\odot$ neutron star. We use the carbon-oxygen core mass at central oxygen depletion as a proxy for the final core mass. We find carbon-oxygen core masses $>2$ M$_\odot$, which is comfortably above the core masses of the aforementioned work (see Extended Data Table~\ref{tab:model_variations_b}). This is indirect evidence that the observed star indeed will reach iron core collapse \cite[see also][]{Vartanyan2021}. 

We assume that the stripped star forms a neutron star with a standard mass of 1.4 M$_\odot$, which means that the ejecta masses should be $\sim$1-2 M$_\odot$. The binaries have a total mass prior to core collapse of 3.8-5.7 M$_\odot$, which means that if we assume only the kick from mass loss is included, and no natal kick is, the binary should form a bound double neutron star system since the ejected mass is less than half of the mass prior to explosion \citep[e.g.,][]{Tauris2023}. 
Before explosion, the relative velocity between the two stars is $v_{\rm rel}\sim 500-1000$ km/s (see Extended Data Table~\ref{tab:model_variations_b}). This is substantially faster than regular natal kicks, suggesting that it would be hard to disrupt the binary even if a natal kick occurs. Models for stripped progenitors suggest that natal kicks could be $\sim 10-500$ km/s \citep{Muller2019}. Observations of Galactic double neutron stars and Be X-ray binaries suggest that the stripped progenitors experienced small to negligible natal kicks \citep[$\lesssim 100$ km/s,][]{Valli2025}. Although the natal kicks from stripped stars are still not entirely understood, we consider that these results point towards small kicks and adopt only the kick from instantaneous mass loss in our predictions.

\paragraph{The double neutron star and GW merger}

Because of the small but non-negligible mass loss during the supernova, we predict moderate eccentricities of 0.3-0.7 for the post-explosion binary systems. The orbital periods also increase and reach a few hours up to almost a day (Extended Data Table~\ref{tab:model_variations_b}). 
Once the binary consists of two compact objects, its orbital evolution is purely driven by gravitational wave radiation, which causes a slow orbital in-spiral. 
We calculate the period and eccentricity evolution due to gravitational wave radiation and the time delay of the eventual merger ($\tau_{\rm delay}$) following \cite{Peters1964}. Specifically, we solve the equations:
\begin{equation}\label{eq:dadt} 
\bigg \langle \dfrac{da}{dt} \bigg\rangle = -\dfrac{64 G^3}{5c^5}\dfrac{M_1M_2M_{\rm tot}}{a^3}\dfrac{1 + \frac{73}{24}e^2 + \frac{37}{96}e^4}{(1-e^2)^{7/2}}
\end{equation}
\begin{equation}\label{eq:dedt} 
\bigg \langle \dfrac{de}{dt} \bigg\rangle = -\dfrac{304}{15}e\dfrac{G^3M_1M_2M_{\rm tot}}{c^5a^4(1-e^2)^{5/2}} (1+\dfrac{121}{304}e^2)
\end{equation}
The evolution of the semi-major axis is then converted to orbital period using Kepler's third law. 
To solve equations \ref{eq:dadt} and \ref{eq:dedt}, we use the \texttt{scipy.integrate.solve\_ivp} solver for ordinary differential equations and adopt its default Runge-Kutta method, which uses a combination of the 4th and 5th order solutions. 
After the supernova, the binary slowly circularizes and tightens. We find that the double neutron stars merge between 160 Myr and 7.9 Gyr after the supernova, which is consistent with the expectations for the double neutron stars in the Galaxy \citep{VignaGomez2018}. Figure~\ref{fig:model} shows a demonstration of how the parameters of the binary change as it evolves, both before and after supernova. 
This suggests that the binary that we describe is the progenitor of a binary neutron star merger detectable via gravitational waves and neutrinos, and potentially as a short gamma-ray burst and a kilonova. 
The resulting merger product could either become a massive neutron star or a low-mass black hole \citep{Sarin2021}.

As described above, whether the system ultimately merges through gravitational-wave emission depends on how much angular momentum is removed from the binary during the non-conservative Case BB mass-transfer episode. If the angular momentum loss is large, like in the case when a circumbinary disk forms (C), the system may merge. However, both the isotropic re-emission and the Jeans loss lead to gravitational wave mergers (A, B, D, E, F, G), also when considering a more massive compact companion (B), a more massive stripped star (F, G), and the remaining hydrogen does not seem to affect the fate of the system either (E). 
While it remains unclear how material leaves a system during mass transfer and how much angular momentum is lost in the process, we note that the standard model (and several other) leads to a binary neutron star merger. Having explored other angular momentum loss schemes, we conclude that the star we present is a very likely GW progenitor. 

\clearpage
\section*{Extended Data}  

\setcounter{figure}{0}
\setcounter{table}{0}
\renewcommand{\figurename}{Extended Data Fig.}
\renewcommand{\tablename}{Extended Data Table}
\renewcommand{\theHfigure}{ED\arabic{figure}}   
\renewcommand{\theHtable}{ED\arabic{table}}
\captionsetup[figure]{labelsep=bar, labelfont=bf, justification=raggedright, singlelinecheck=false}

\begin{table*}[p]
\centering
\caption{Radial velocity measurements shown in \cref{fig:ellcjoker}}.
\label{tab:RV_measurements}
\begin{minipage}[b]{0.45\textwidth}\centering
\begin{tabular}{lcc}
\toprule 
Date & Radial Velocity & Error \\ 
\midrule 
\footnotesize{[HJD]} & 
\footnotesize{[km/s]} & 
\footnotesize{[km/s]} \\ 
\midrule 
2458838.5585 & 191.41 & 10.70 \\
2458838.5727 & 116.25 & 8.35 \\
2458838.5868 & 51.06 & 12.93 \\
2458839.6649 & 59.04 & 13.59 \\
2458840.6321 & 244.4 & 11.15 \\
2458866.5597 & 63.27 & 10.99 \\
2458866.5739 & 38.03 & 8.31 \\
2458866.588 & 54.27 & 10.20 \\
2459202.6068 & 252.86 & 11.35 \\
2459202.621 & 355.34 & 10.23 \\
2459202.657 & 275.36 & 12.66 \\
2459202.6712 & 207.62 & 10.31 \\
2459203.5875 & 21.43 & 9.34 \\
2459203.6016 & 36.77 & 12.08 \\
2459203.7034 & 282.01 & 9.09 \\
2459204.7038 & 294.93 & 14.05 \\
2459204.7179 & 307.73 & 15.31 \\
2459246.5551 & 291.39 & 13.53 \\
\vdots & \vdots & \vdots  \\
\bottomrule 
\end{tabular}
\end{minipage} 
\hspace{0.5cm}
\begin{minipage}[b]{0.45\textwidth}\centering
\begin{tabular}{lcc}
\toprule 
Date & Radial Velocity & Error \\ 
\midrule 
\footnotesize{[HJD]} & 
\footnotesize{[km/s]} & 
\footnotesize{[km/s]} \\ 
\midrule 
\vdots & \vdots & \vdots  \\
2459246.5693 & 313.62 & 13.06 \\
2459249.5394 & 345.67 & 15.12 \\
2459249.5535 & 299.62 & 12.12 \\
2459249.5676 & 266.37 & 10.11 \\
2459251.5672 & 112.19 & 11.08 \\
2459251.5814 & 163.55 & 10.09 \\
2459575.6318 & 282.71 & 9.28 \\
2459575.6459 & 236.11 & 11.20 \\
2459586.5484 & 288.0 & 14.81 \\
2459587.5936 & 297.71 & 10.71 \\
2459865.7927 & 72.52 & 15.91 \\
2459865.8068 & 20.28 & 14.78 \\
2459869.7118 & 313.27 & 8.07 \\
2459869.726 & 355.48 & 8.01 \\
2459869.7401 & 301.36 & 9.42 \\
2459869.7554 & 263.24 & 9.86 \\
2459869.7696 & 176.03 & 9.49 \\
2459869.7837 & 91.58 & 9.32 \\
\bottomrule 
\end{tabular}
\end{minipage}
\end{table*}

\begin{table*}[p]
\begin{center}
\caption{Model variations: initial parameters and case BB mass transfer. We initiate models A-D with the same stripped star model, while model E contains a pure helium star and model F a more massive stripped star. All models are initiated with a circular 4.17 hour orbit. 
From left to right we list the model name, the initial mass of the stripped star, the initial mass of the compact companion, the adopted treatment of angular momentum loss during case BB mass transfer, the amount of material accreted onto the compact object, and the spin period resulting from case BB mass transfer and assuming the compact object is a neutron star. Core-collapse and post-collapse outcomes for these same models are given in Extended Data Table~\ref{tab:model_variations_b}. 
}
\label{tab:model_variations_a}
\vspace{1mm}
{\scriptsize
\begin{tabular}{lccccc}
\toprule \midrule 
Model & $M_{\rm strip\, init}$& $M_{\rm 2\, init}$ & Angular Momentum Loss & $\Delta M_{\rm 2\, acc}$ & $P_{\rm 2\, spin}$ \\ 
& $[M_\odot]$ & $[M_\odot]$ &  & $[10^{-4} M_\odot]$ & [ms] \\ 
\midrule 
A & 3.61 & 1.4 & Isotropic re-emission & 7.2 & 79 \\ 
B & 3.61 & 3 & Isotropic re-emission & -- & -- \\ 
C & 3.61 & 1.4 & Circumbinary disk & -- & -- \\ 
D & 3.61 & 1.4 & Jeans loss & -- & -- \\  
E & 3.60 & 1.4 & Isotropic re-emission & 4.2 & 118 \\ 
F & 4.28 & 1.4 & Isotropic re-emission & 4.2 & 119 \\ 
G & 4.98 & 1.4 & Isotropic re-emission & 2.4 & 183 \\ 
\bottomrule 
\end{tabular}
}
\end{center}
\textbf{Notes.} We consider accretion to be fully non-conservative in model B because of the more massive companion and in model D simply because the Eddington-limited accretion scheme in MESA is limited to isotropic re-emission. Model C results in a merger during the Case BB interaction and therefore does not contain information about the late evolutionary stages. 
\end{table*}

\begin{table*}[p]
\begin{center}
\caption{Model variations: properties of the binary immediately before and after core collapse, continued from Extended Data Table~\ref{tab:model_variations_a}. From left to right we list the model name, the stripped star's mass, helium mass, and carbon-oxygen core mass at core collapse, the pre-supernova relative velocity between the two binary components, the ejecta mass from the supernova assuming the stripped star forms a 1.4 M$_\odot$ NS, the pre-supernova orbital period, the post-supernova orbital period and eccentricity, and the time delay between core collapse and gravitational wave merger. 
}
\label{tab:model_variations_b}
\vspace{1mm}
{\scriptsize
\begin{tabular}{lcccccccccc}
\toprule \midrule 
& \multicolumn{6}{c}{Pre Core Collapse} & \multicolumn{3}{c}{Post Core Collapse} \\
\cmidrule(lr){2-7} \cmidrule(lr){8-10}
Model & $M_{\rm strip}^{\rm total}$ & He mass & CO core mass & $v_{\rm rel}$ & $M_{\rm ejecta}$ & $P_{\rm i}$ & $P_{\rm f}$ & $e_{\rm f}$ & $\tau_{\rm delay}$ \\ [3pt]
& $[M_\odot]$& $[M_\odot]$ & $[M_\odot]$ & [km/s] & $[M_\odot]$ & [hours] & [hours] && [Myr] \\ 
\midrule 
A & 2.44 & 0.22 & 2.13 & 777 & 1.04 & 1.89 & 4.43 & 0.37 & 220 \\ 
B & 2.66 & 0.42 & 2.11 & 619 & 1.26 & 5.51 & 10.3 & 0.29 & 1432 \\ 
C & -- & -- & -- & -- & -- & -- & -- & -- & -- \\ 
D & 2.75 & 0.52 & 2.16 & 541 & 1.35 & 6.07 & 19.8 & 0.48 & 7883 \\ 
E & 2.44 & 0.21 & 2.15 & 774 & 1.04 & 1.92 & 4.50 & 0.37 & 229 \\ 
F & 3.08 & 0.34 & 2.65 & 905 & 1.68 & 1.40 & 7.05 & 0.60 & 264 \\ 
G & 3.55 & 0.22 & 3.22 & 1086 & 2.15 & 0.90 & 10.6 & 0.77 & 175 \\ 
\bottomrule 
\end{tabular}
}
\end{center}
\textbf{Notes.} All quantities in this table describe the stripped star and its orbit; the compact companion is assumed unaffected by the explosion aside from its orbital motion. Model C results in a merger during the Case BB interaction and therefore does not contain information about the late evolutionary stages. 
\end{table*}

\clearpage
\begin{figure*}[p]
    \centering
    \includegraphics[width=0.3\textwidth]{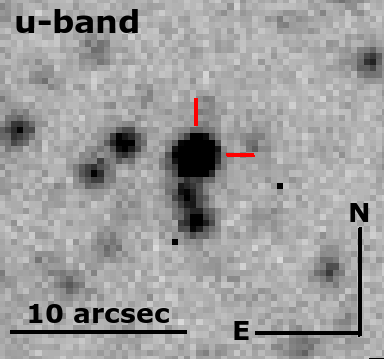}
    \includegraphics[width=0.3\textwidth]{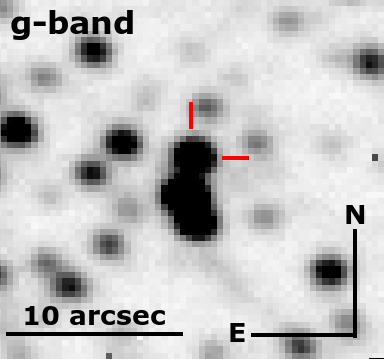}
    \includegraphics[width=0.3\textwidth]{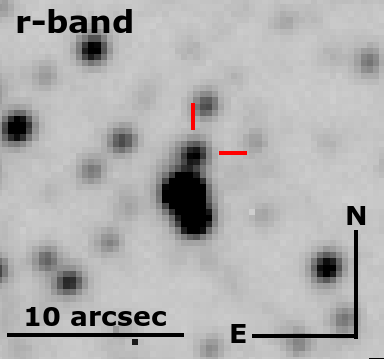}
    \caption{\textbf{ULTRACAM images of the 20$\times$20$\arcsec$ region surrounding \name{} in the Magellanic Clouds.} \textbf{a, b, c,}  observations obtained in the u, g, and r-bands, respectively. \name{} is indicated by the red cross hairs in each panel. When performing photometry, we simultaneously fit \name{} along with the two other stars immediately to the south. }
    \label{fig:image-cutouts}
\end{figure*}

\begin{figure}[p]
    \centering
    \includegraphics[width=\textwidth]{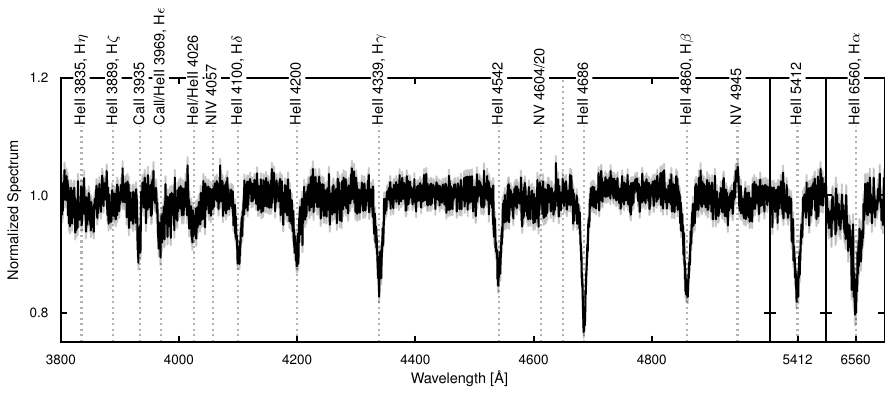}
    \caption{\textbf{High-S/N stacked optical spectrum of \name.} Epoch spectra shifted to the rest frame of the stripped star using the measured RVs and co-added following \citep{Laroche2026}. The black line shows the continuum normalized stacked flux with gray shading showing its 1$\sigma$ uncertainty. Dotted vertical lines mark the rest wavelengths of the Balmer, He~\textsc{ii}, Ca~\textsc{ii}, He~\textsc{i}, N~\textsc{iv} and N~\textsc{v} lines labeled at the top. The strong absorption lines are photospheric and arise from the stripped star, and each Balmer line is blended with a He~\textsc{ii} Pickering ($n\to4$) line. The wavelength axis is broken to show the He~\textsc{ii}\,5412 and H$\alpha$ regions.
    }
    \label{fig:spectrum}
\end{figure}

\begin{figure}[p]
    \centering
    \includegraphics[]{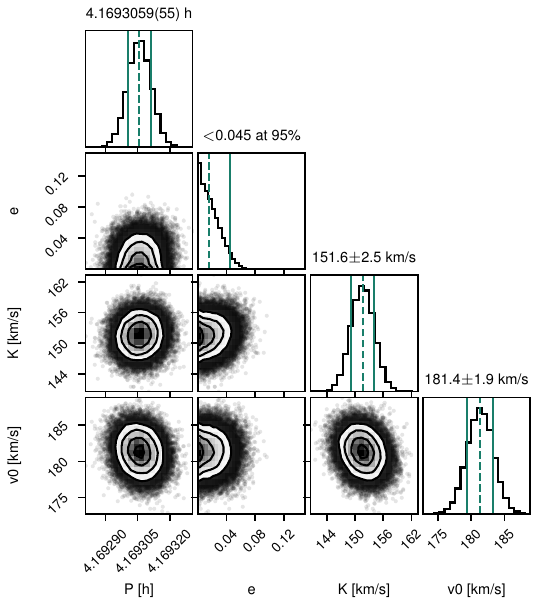}
    \caption{\textbf{Corner plot for Eccentric Orbit Solution}. Contours and vertical lines for $P$, $K$, and $v_0$ mark the 16th, 50th, and 84th percentiles ($1\sigma$ equivalent). For eccentricity we show the median and the 95\% credible upper limit (95th percentile of the posterior). The corner plot for the circular solution is in Extended Data \cref{fig:circularcorner}.}
    \label{fig:eccentriccorner}
\end{figure}

\begin{figure}[p]
    \centering
    \includegraphics[]{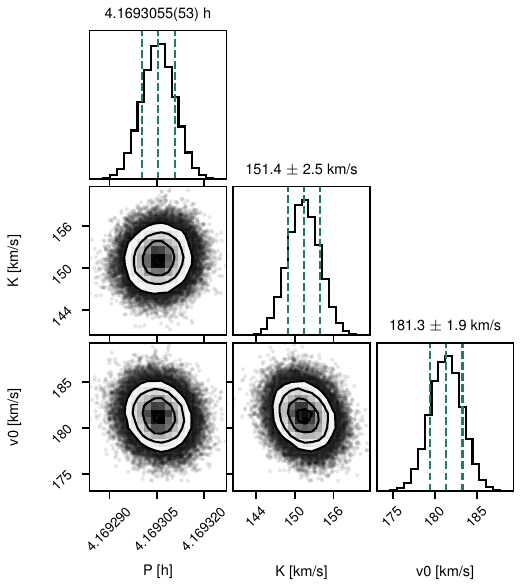}
    \caption{\textbf{Corner plot for Circular Orbit Solution}. Same as Extended Data \cref{fig:eccentriccorner} but for the circular solution.}
    \label{fig:circularcorner}
\end{figure}

\begin{figure*}[p]
\centering
\includegraphics[]{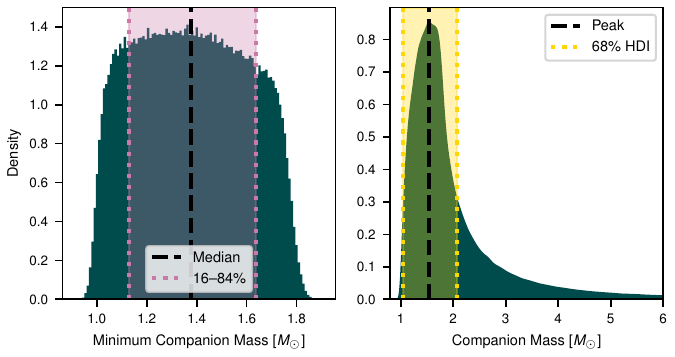}
\caption{\textbf{Companion mass for edge-on and isotropic inclinations.} \textbf{a}, Minimum companion mass for an edge-on orbit (i=90\degree), from the binary mass function and the evolutionary stripped-star mass. The dashed black line is the median (\MCompanionOrbitMedian\,\Msun); the shaded pink band is the equal-tailed 68\% interval [\MCompanionOrbitPercentiles\,\Msun]. \textbf{b}, Companion mass for the same draws of stripped-star mass and mass function, with inclination sampled from an isotropic distribution, $P(i)\propto\sin i$. The dashed black line is the peak of the smoothed distribution (\MCompanionOrbitDistPeak\,\Msun); the shaded gold band is the 68\% HDI [\MCompanionOrbitDistRange\,\Msun].}

    \label{fig:minmass}
\end{figure*}

\begin{figure}[p] 
    \centering
    \includegraphics{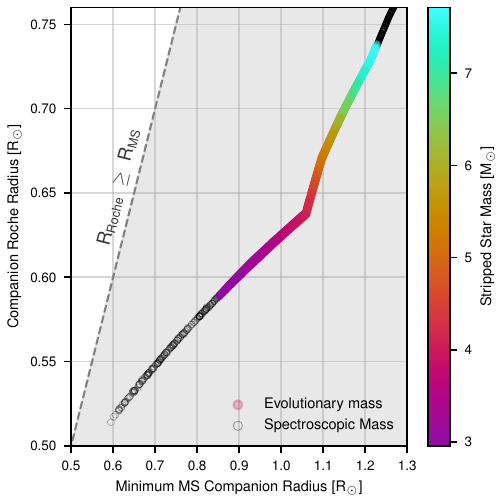}
    \caption{\textbf{The size of the orbit rules out a \gls{MS}  star}. For each stripped star mass sampled from the evolutionary (colored points) and spectroscopic (grey points) mass distributions, we compute the minimum companion mass implied by the binary mass function, and the corresponding minimum radius from MIST models for a \gls{MS} companion. This is compared to the Roche lobe radius set by the observed orbital configuration; the dashed line marks equality between the two. In the shaded region, a \gls{MS} companion would overfill its Roche lobe, making it incompatible with the observed orbit.}
    \label{fig:noms}
\end{figure}

\begin{figure*}[p] 
    \centering
    \includegraphics[width=0.6\columnwidth]{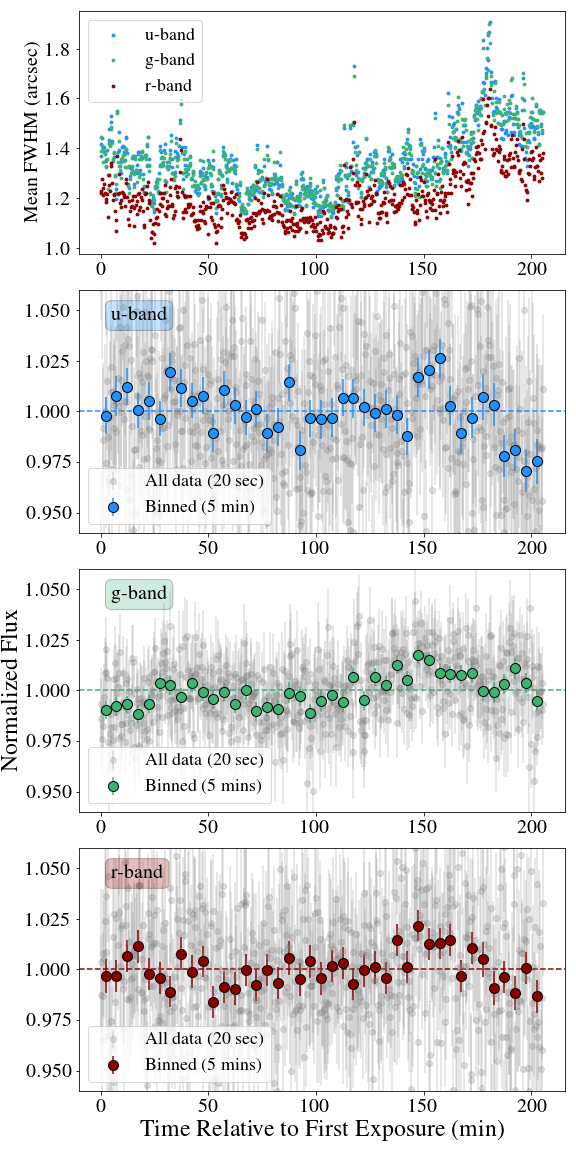}
    \caption{\textbf{Summary of ULTRACAM light curves for \name{}.} \textbf{b}, \textbf{c}, and \textbf{d} show the u, g, and r-band light curves, respectively. Grey points show results for individual images (taken every $\sim$20 seconds), while colored points represent the results from binning 15 sequential images (corresponding to $\sim$5 minutes of total exposure). Each panel is normalized to the mean flux in that band. Small variations on the order of $\lesssim$2.5\% are seen in all bands, although the photometric errors are smallest in the $g$-band. \textbf{a} shows the mean FWHM for the PSF measured in each image. The seeing varied from $\sim1.1-1.6$$\arcsec$ throughout the observations. However, the observed variability does not directly trace these changes. }
    \label{fig:uc-all}
\end{figure*}

\begin{figure}[p]
\includegraphics[width=\textwidth]{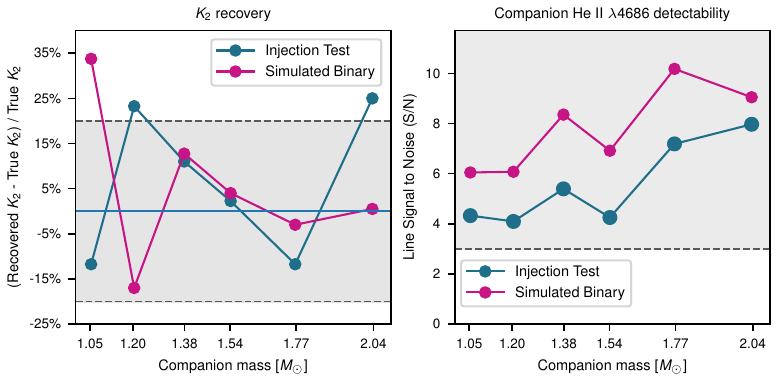}
\caption{\textbf{Spectral disentangling excludes a companion stripped star.}
\textbf{a}, Percent difference between the true $K_2$ and the $K_2$ recovered after disentangling. The shaded region corresponds to $\pm 20\%$ for which $K_2$ is typically recovered. \textbf{b}, He\,II~$\lambda$4686 line S/N in the
disentangled secondary spectrum for the same runs; the dashed line at
$\mathrm{S/N} = 3$ is our nominal detection threshold.}
\label{fig:k2-recovery-snr}
\end{figure}

\begin{figure*}[p]
\includegraphics[width=\textwidth]{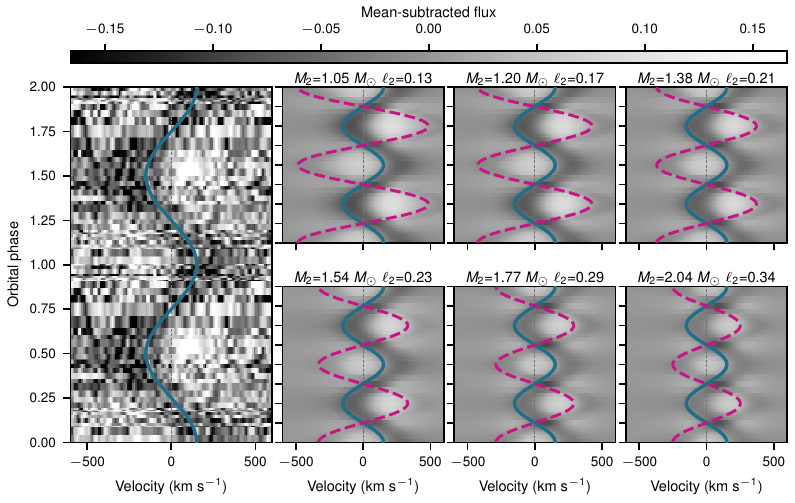}
\caption{\textbf{Observed dynamical RV compared to models including a stripped star companion.}
\textbf{a}, MagE mean-subtracted dynamical residual as a function of
heliocentric velocity and orbital phase (two cycles). \textbf{b}, Simulated binaries with a \name\ primary and stripped star companions for $M_2 = 1.05$--$2.04$\Msun\ on the same grid. Curves mark the RV curve for the primary (teal) and the companion (magenta) modeled spectra.}
\label{fig:dynamical-4686}
\end{figure*}

\begin{figure}[p]
\centering
\includegraphics[width=0.49\textwidth]{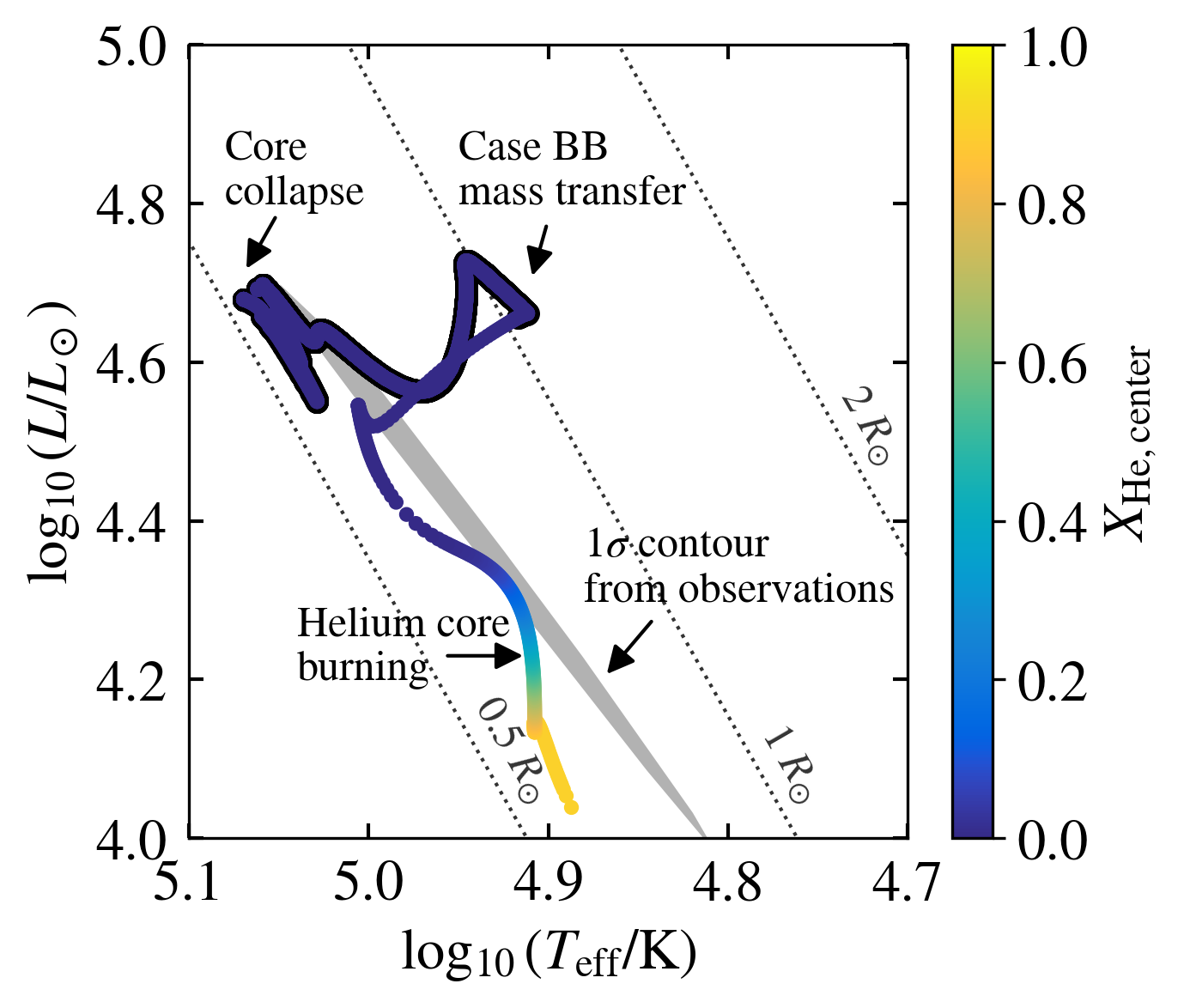}
\includegraphics[width=0.49\textwidth]{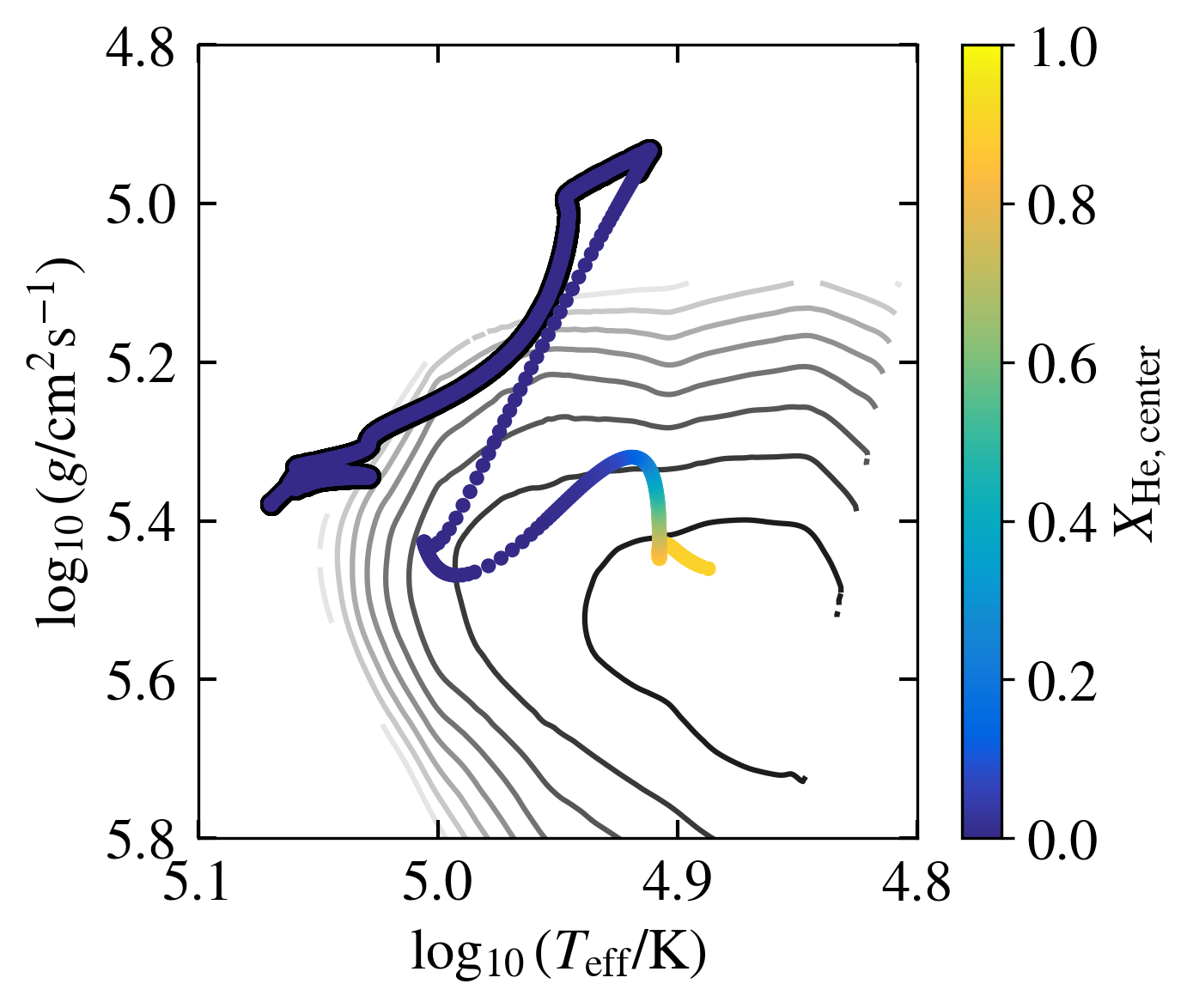}
\caption{\textbf{Modeling of the future evolution of the stripped star binary in \name.} Here, we display our fiducial evolutionary model (A in Tables~\ref{tab:model_variations_a} and \ref{tab:model_variations_b}), which has 3.6 M$_\odot$ and is stripped down to the surface hydrogen mass fraction ($X_{\rm H,\, surface}$) of 0.2. We consider that this model matches the observational constraints best. \textbf{a}, a Hertzsprung-Russell diagram with the spectral models allowed within 1$\sigma$ from \cite{Gotberg2023} marked as a gray area and the evolutionary model we present here as a colored track. Color represent central helium mass fraction ($X_{\rm He,\,center}$), thus highlighting the long-lasting central helium burning phase through color change. The case BB mass transfer is marked using black background for the track, and the track ends at central oxygen depletion, which we label ``core collapse." The figure shows overlap between the evolutionary model during central helium burning and the 1$\sigma$ contour. 
\textbf{b}, a Kiel diagram where we compare the evolutionary model with the spectral models allowed within $1\sigma$. The contours outline the spectral models allowed within 0.2, 0.3, ... 0.9$\sigma$. The helium-core burning phase of the evolutionary model falls well within the contours. 
In the two panels, we have plotted the spectral models that have a $X_{\rm H, \, surface} = 0.2$ and that are allowed within 1$\sigma$ from the spectroscopic fit. 
}
\label{fig:stripped_model}
\end{figure}

\begin{figure}[p]
\centering
\includegraphics[width=0.5\textwidth]{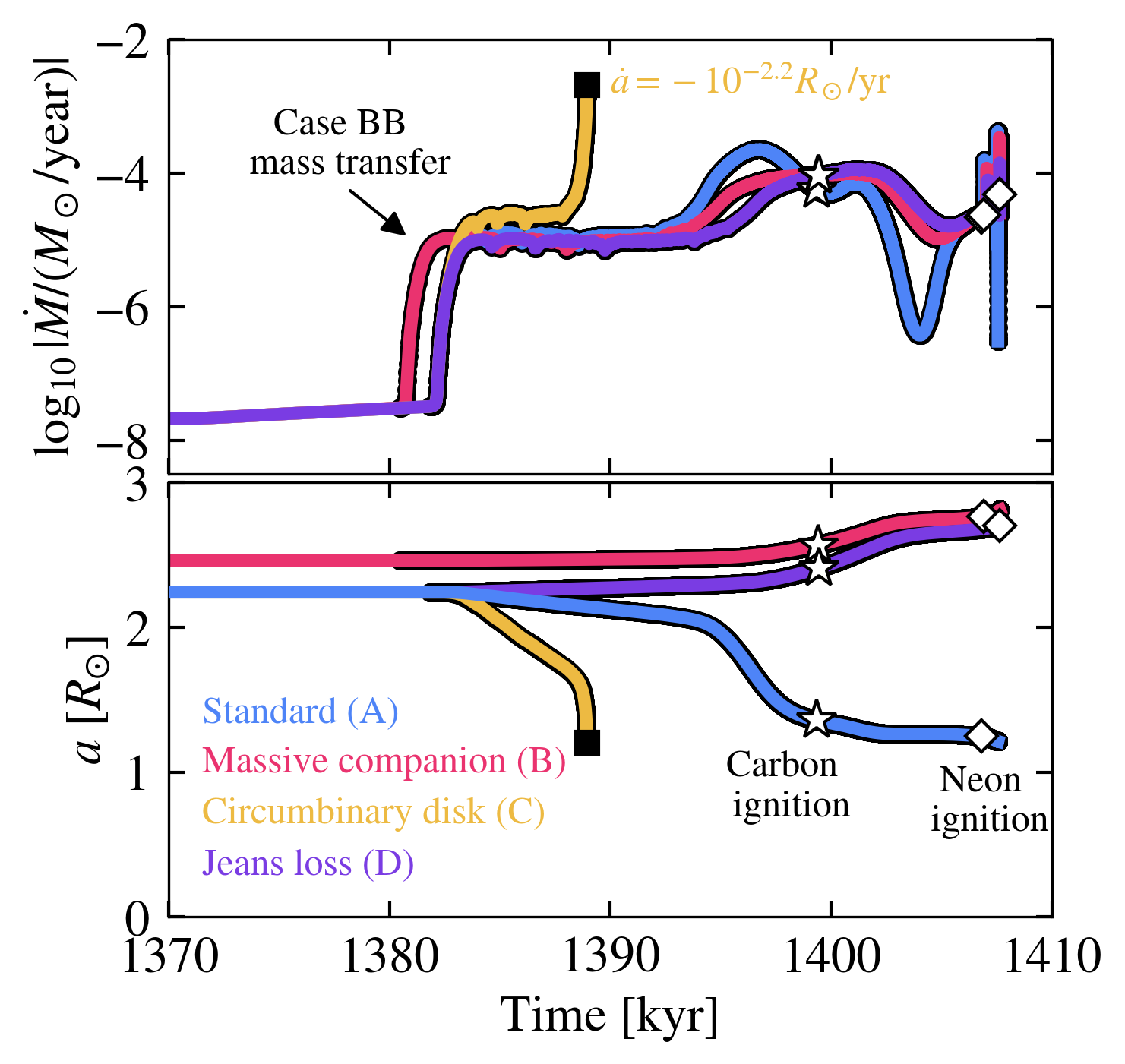}
\caption{\textbf{Orbital evolution variations.} Evolution of the mass-loss rate (top) and orbital separation (bottom) for models A, B, C, and D, which have the same initial stripped star model. While in the standard model (A), the orbit tightens during Case BB mass transfer (labeled and marked with black), the circumbinary disk model (C) includes such rapid drainage of angular momentum that the orbital shrinkage cannot be maintained for long and the model stops, when the orbital shrinkage reaches $-10^{-2.2}$ R$_\odot$/year (marked with a black box). Both the model with a massive companion (B) and the angular momentum lost from the donor (D) lead to orbital widening. We mark central ignition of carbon and neon using star and diamond shaped markers, respectively.}
\label{fig:Mdot_a}
\end{figure}

\begin{figure}[p]
\centering
\includegraphics[width=0.5\textwidth]{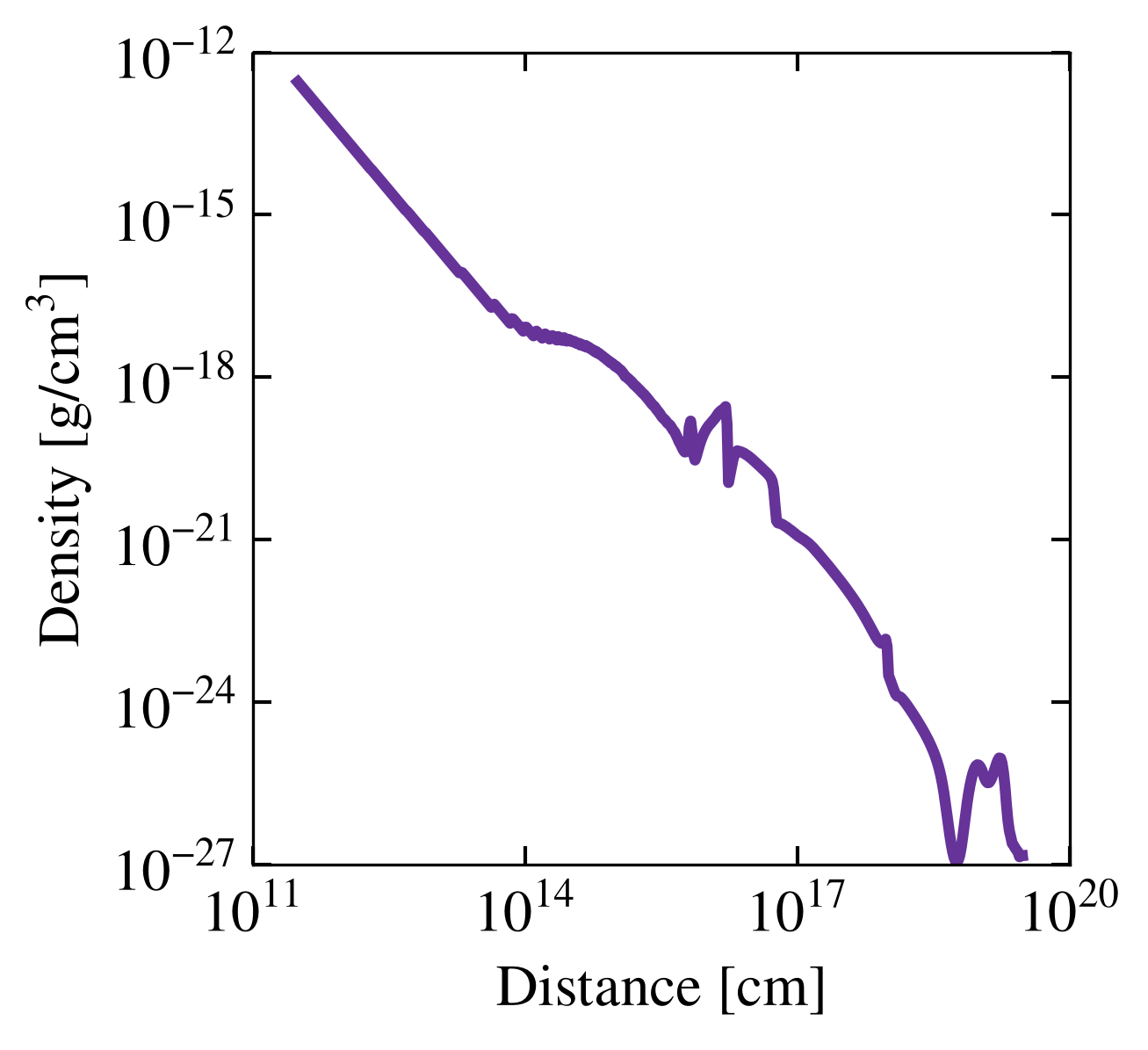}
\caption{\textbf{Expected circumstellar density structure at core collapse.} We plot the density of the circumstellar material ejected from the non-conservative mass transfer in model A as function of radius in centimeters and at core collapse.}
\label{fig:csm_density}
\end{figure}

\clearpage
\pagebreak
\bibliography{references}

\end{document}